\documentclass[useAMS,usenatbib]{mn2e}

\usepackage{pdflscape}
\usepackage{graphicx}

\def\teff{T_{\rm eff}}
\def\feh{\rm[Fe/H]}
\def\logg{\log\,g}

\def\afe{\rm[\alpha/Fe]}
\def\msun{\rm M_{\odot}}
\def\runo{r_{010}}
\def\rdos{r_{02}}
\def\dnu{\Delta\nu}
\def\num{\nu_\mathrm{max}}
\def\CM{{\cal M}}
\def\CP{{\cal P}}

\def\CG{{\cal G}}
\def\CH{{\cal H}}

\def\nuref{\nu_{nl}^{\rm (ref)}}

\begin{document}

\title[Ages of exoplanet host stars]{Ages and fundamental properties of {\it Kepler} exoplanet host stars from asteroseismology}
\author[V.~Silva Aguirre et al.]{V.~Silva Aguirre$^1$, G.~R. Davies$^{2,1}$, S.~Basu$^3$, J.~Christensen-Dalsgaard$^1$, O.~Creevey$^4$,\newauthor 
T.~S. Metcalfe$^{5,1}$, T.~R. Bedding$^{6,1}$, L.~Casagrande$^7$, R. Handberg$^{1,2}$, M.~N. Lund$^{1,2}$,\newauthor P.~E. Nissen$^1$, W.~J. Chaplin$^{2,1}$, D. Huber$^{6,8,1}$, A.~M. Serenelli$^9$, D. Stello$^{6,1}$,\newauthor V. Van Eylen$^1$, T.~L. Campante$^{2,1}$, Y. Elsworth$^{2,1}$, R.~L. Gilliland$^{10}$, S. Hekker$^{11,1}$,\newauthor C. Karoff$^{12,1}$, S.~D. Kawaler$^{13}$, H. Kjeldsen$^{1}$, M.~S. Lundkvist$^{1}$ \\
$^1$Stellar Astrophysics Centre, Department of Physics and Astronomy, Aarhus University, Ny Munkegade 120, DK-8000 Aarhus C, Denmark\\
$^2$School of Physics and Astronomy, University of Birmingham, Birmingham, B15 2TT, UK \\
$^3$Department of Astronomy, Yale University, PO Box 208101, New Haven, CT 06520-8101, USA\\
$^4$Institut dÕAstrophysique Spatiale, Universit\'e Paris XI, UMR 8617, CNRS, Batiment 121, F-91405 Orsay Cedex, France\\
$^5$Space Science Institute, 4750 Walnut St. Suite 205, Boulder CO 80301 USA\\
$^6$Sydney Institute for Astronomy (SIfA), School of Physics, University of Sydney, NSW 2006, Australia\\
$^7$Research School of Astronomy \& Astrophysics, Mount Stromlo Observatory, The Australian National University, ACT 2611, Australia\\
$^8$SETI Institute, 189 Bernardo Avenue, Mountain View, CA 94043, USA\\
$^9$Instituto de Ciencias del Espacio (ICE-CSIC/IEEC), Campus UAB, Carrer de Can Magrans, s/n, 08193 Cerdanyola del Valles, Spain\\
$^{10}$Center for Exoplanets and Habitable Worlds, The Pennsylvania State University, University Park, PA 16802, USA\\
$^{11}$Max-Planck-Institut f\"ur Sonnensystemforschung, Justus-von-Liebig-Weg 3, 37077 G\"ottingen, Germany\\
$^{12}$Department of Geoscience, Aarhus University, H¿egh-Guldbergs Gade 2, 8000, Aarhus C, Denmark\\
$^{13}$Department of Physics and Astronomy, Iowa State University, Ames, IA 50014, USA}
\maketitle
\begin{abstract}
We present a study of 33 {\it Kepler} planet-candidate host stars for which asteroseismic observations have sufficiently high signal-to-noise ratio to allow extraction of individual pulsation frequencies. We implement a new Bayesian scheme that is flexible in its input to process individual oscillation frequencies, combinations of them, and average asteroseismic parameters, and derive robust fundamental properties for these targets. Applying this scheme to grids of evolutionary models yields stellar properties with median statistical uncertainties of 1.2\% (radius), 1.7\% (density), 3.3\% (mass), 4.4\% (distance), and 14\% (age), making this the exoplanet host-star sample with the most precise and uniformly determined fundamental parameters to date. We assess the systematics from changes in the solar abundances and mixing-length parameter, showing that they are smaller than the statistical errors. We also determine the stellar properties with three other fitting algorithms and explore the systematics arising from using different evolution and pulsation codes, resulting in 1\% in density and radius, and 2\% and 7\% in mass and age, respectively. We confirm previous findings of the initial helium abundance being a source of systematics comparable to our statistical uncertainties, and discuss future prospects for constraining this parameter by combining asteroseismology and data from space missions. Finally we compare our derived properties with those obtained using the global average asteroseismic observables along with effective temperature and metallicity, finding an excellent level of agreement. Owing to selection effects, our results show that the majority of the high signal-to-noise ratio asteroseismic {\it Kepler} host stars are older than the Sun.
\end{abstract}

\section{Introduction}\label{sec-int}
The CoRoT and {\it Kepler} missions have revolutionised the quality and quantity of data that are available for the analysis of solar-like oscillations \citep[see e.g.,][]{Chaplin:2013gz}. As a result, asteroseismology of solar-type stars has matured into a powerful tool to help characterise extrasolar planetary systems. Even a relatively straightforward analysis of solar-like oscillations can usually provide the surface gravity and mean density of an exoplanet host star with a typical precision much higher than the constraints available from transit modelling \citep[e.g.,][Van Eylen et al. 2015, submitted]{Huber:2013jb}. With an independent effective temperature constraint, a precise stellar radius from asteroseismology can be obtained and used to determine the absolute planetary radius from transit photometry. The asteroseismic mass provides the absolute scale of the orbit, and when combined with transit-timing or radial velocity measurements yields the absolute planetary mass \citep{Carter:2012gq,Marcy:2014hr}. The asteroseismic age can be used to assess the dynamical stability of the system, and to establish its chronology with respect to other planetary systems. For the best and brightest targets, asteroseismology can also provide an independent determination of the stellar rotation rate and inclination \citep{Chaplin:2013dg,VanEylen:2014cy,Davies:2015dl}. This can be used to probe spin-orbit alignment for systems without Rossiter-McLaughlin (R-M) measurements during transit, and it provides the full three-dimensional orientation when R-M measurements are available \citep[e.g.,][]{Benomar:2014eo,Lund:2014ez}. As a further matter, lower limits on the surface gravities of exoplanet-host stars can still be placed in the event of a non-detection of solar-like oscillations \citep[][]{Campante:2014hr}.

Prior to the CoRoT and {\it Kepler} missions, solar-like oscillations were extremely difficult to observe for anything but the Sun, and data only existed for a handful of the brightest stars in the sky \citep[e.g.,][]{1995AJ....109.1313K,Bouchy:2001ca,Carrier:2005gu,Arentoft:2008fn}. With the advent of space-born asteroseismic observations, thousands of detections have been possible for stars in different evolutionary stages \citep[e.g.,][]{DeRidder:2009cd,Huber:2011be}. The focus of recent efforts has been to develop uniform data analysis and modelling strategies \citep{Appourchaux:2012kd, Chaplin:2014jf}, and to optimise the precision of asteroseismic inferences by matching the individual oscillation frequencies or ratios of characteristic frequency separations \citep{SilvaAguirre:2013in,Metcalfe:2014ig}. These techniques typically improve the precision by a factor of two or more over methods that only use the global oscillation parameters \citep[e.g.,][]{Mathur:2012bj,SilvaAguirre:2013in,Lebreton:2014gf}, but the absolute accuracy is difficult to assess. Initial comparisons for the small samples of dwarfs and subgiants with well-determined distances and interferometric observations suggest consistency at the level of 4--5\% in stellar radii (\citet{SilvaAguirre:2012du, Huber:2012iv}, see also \citet{Miglio:2013fb} for a summary).

Future space-based photometry and astrometry, as well as complementary data from ground-based networks, will soon yield improved asteroseismic constraints for many exoplanet host stars. Observations of bright stars and clusters in the ecliptic plane have already started with K2 \citep{Howell:2014ju}, and the \emph{TESS} mission will soon obtain similar data over the entire sky \citep{2014SPIE.9143E..20R}. At the same time, the \emph{Gaia} mission will determine improved parallaxes for all of these targets \citep{Perryman:2001cp}, yielding luminosity constraints that could help break intrinsic parameter correlations in stellar models. The \emph{PLATO} mission will build upon these successes with extended time-series of many fields around the Galactic plane \citep{Rauer:2014kx}, while ground-based spectroscopy from the Stellar Observations Network Group \citep[SONG,][]{2008CoAst.157..273G} will complement the satellite data for the brightest stars in the sky. Now is the time to address the underlying sources of systematic uncertainty in stellar properties determination, before this next deluge of observations. Exploration of biases in asteroseismic models and fitting methods for individual targets has been made \citep[e.g.,][]{Miglio:2005is,Lebreton:2014gf}, and we intend in this study to complement these efforts using a larger sample of stars together with a range of evolutionary and pulsation codes, and fitting algorithms.

In an accompanying paper (Davies et al. 2015, in preparation) the extraction of individual frequencies, combinations, and derivation of correlations has been performed for a sample of \emph{Kepler} Objects of Interest (KOIs) with high-quality asteroseismic observations. We make use of these data to determine a robust set of fundamental properties, including ages, by introducing a new BAyesian STellar Algorithm (BASTA) and grids of evolutionary tracks specially constructed for this purpose. Our modelling strategy initially adopts this fitting algorithm and applies it to a fixed set of observables and modelling ingredients, and uses the results as the comparison basis for quantifying the impact of additional input physics (e.g.\ diffusion, convective overshoot, mixing-length, composition) in a systematic manner. We further explore the effects introduced by choosing different sets of asteroseismic observables, evolutionary and pulsation codes, as well as treatment of the pulsation data by determining stellar properties with three asteroseismic pipelines widely used in the literature.

This paper is organised as follows. In Section~\ref{sec_targ} we describe the target sample and define the observational quantities to be matched, while our new model grids and Bayesian analysis scheme are outlined in Section~\ref{sec_par}. The main results are discussed in Section~\ref{sec_res}, including a study of systematic uncertainties introduced by changes in the input physics and a thorough comparison with the results of asteroseismic fitting algorithms described in Appendix~\ref{app_meth}. We use our derived stellar properties to check for consistency with the results from empirical scaling relations in Section~\ref{sec_scal}, and explore possible correlations of planetary properties with age in Section~\ref{sec_exo}. Conclusions and closing remarks are given in Section~\ref{sec_conc}.
\section{Target selection and data}\label{sec_targ}
\begin{figure*}
\includegraphics[width=84mm]{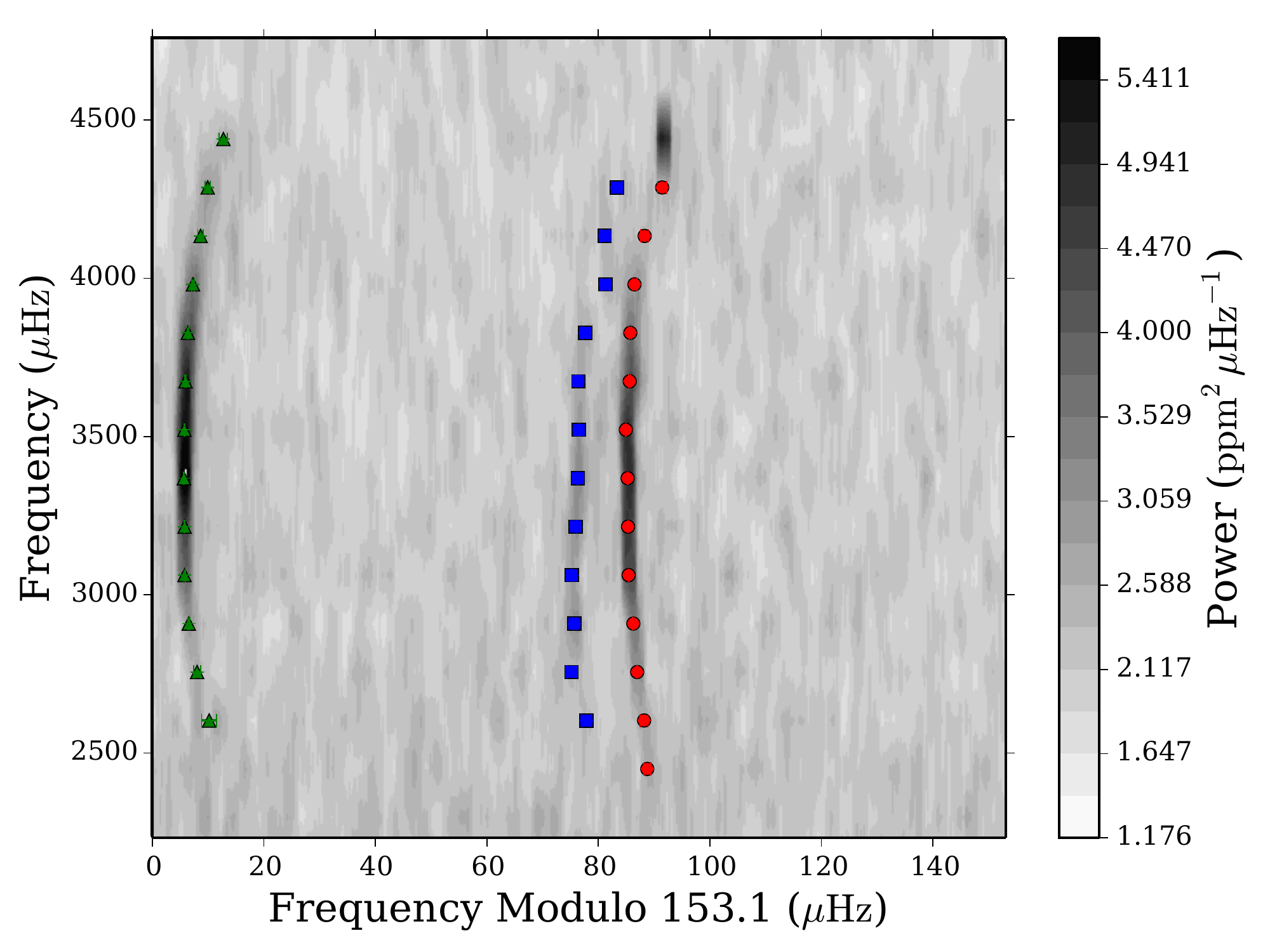}
\includegraphics[width=84mm]{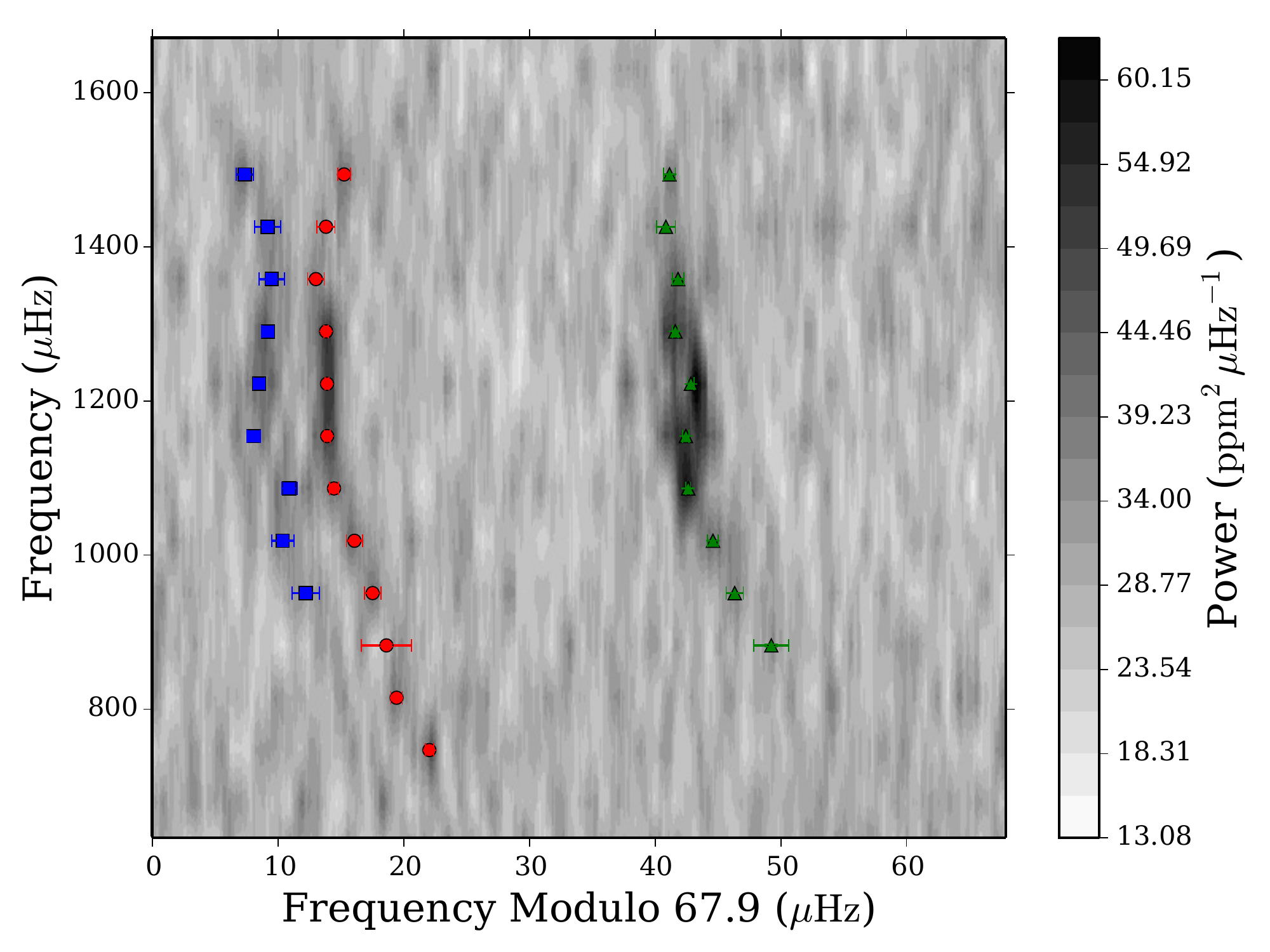}
\caption{\'Echelle diagram of two selected targets from the sample. Symbols depict the different angular-degree values $\ell=0$ (circles), $\ell=1$ (triangles), and $\ell=2$ (squares). Uncertainties in some frequencies are smaller than symbol sizes. {\it Left}: Kepler-409 shows the quasi-vertical alignment of its ridges as expected from pure p-mode frequencies. {\it Right}: Kepler-36 presents signature of modes with mixed character in the $\ell=2$ ridge at $\sim$1070~$\mu$Hz.}
\label{fig:ech_mix}
\end{figure*}

Our sample has been extracted from the 77 exoplanet host stars presented in \citet{Huber:2013jb}. These stars show clear signatures of stochastically excited oscillations with power spectra characterised by a gaussian-shaped envelope centred on the frequency of maximum power $\num$. Pulsation modes of the same angular degree $\ell$ and consecutive radial order $n$ are separated in frequency by a roughly constant spacing, known as the large frequency separation, defined as
\begin{equation}\label{eqn:dnu}
\Delta\nu_{\ell}(n)=\nu_{n,\ell}-\nu_{n-1,\ell}\,.
\end{equation}

\citet{Huber:2013jb} determined (for all 77 targets) the effective temperature $\teff$ and surface composition $\feh$ from high-resolution spectroscopy, as well as the global average seismic parameters $\langle\dnu\rangle$ and $\num$. From this sample, 35 stars have observations with high enough signal-to-noise-ratios (SNRs) for reliable individual frequencies to be extracted (see companion paper by Davies et al. 2015, in preparation). Typically, more than six consecutive modes are detected for each angular degree $\ell=0-2$ allowing the construction of frequency combinations that aim at isolating the contribution of the inner structure from that of the outer layers. In particular, the frequency ratios $\rdos$, $r_{01}$, and $r_{10}$ defined as \citep[][]{Roxburgh:2003bb}
\begin{equation}\label{eqn:r02}
r_{02}(n)=\frac{d_{02}(n)}{\Delta\nu_{1}(n)}
\end{equation}
\begin{eqnarray}\label{eqn:rat}
r_{01}(n)=\frac{d_{01}(n)}{\Delta\nu_{1}(n)},& & r_{10} = \frac{d_{10}(n)}{\Delta\nu_{0}(n+1)}\,,
\end{eqnarray}
where $d_{02}(n)=\nu_{n,0}-\nu_{n-1,2}$ is the small frequency separation and $d_{01}(n)$ and $d_{10}(n)$ are the smooth 5-point small frequency separations:
\begin{equation}\label{eqn:d01}
d_{01}(n)=\frac{1}{8}(\nu_{n-1,0}-4\nu_{n-1,1}+6\nu_{n,0}-4\nu_{n,1}+\nu_{n+1,0})
\end{equation}
\begin{equation}\label{eqn:d10}
d_{10}(n)=-\frac{1}{8}(\nu_{n-1,1}-4\nu_{n,0}+6\nu_{n,1}-4\nu_{n+1,0}+\nu_{n+1,1})\,.
\end{equation}

The methods considered in this paper to determine stellar properties use different sets of asteroseismic observables in their optimisation procedure (for details see sections~\ref{ssec_bayes} and~\ref{ssec_meth} below, as well as Appendix~\ref{app_meth}): only the individual oscillation frequencies, only the frequency ratios, or a weighted average of both sets. For reference, typical uncertainties in the three closest frequencies to $\num$ are $\sim$0.3~$\mu$Hz (and below 0.1~$\mu$Hz in the best cases), while fractional uncertainties in the same range for the frequency ratios are $\sim$7\% (and smaller than 3\% in the highest SNR stars, further details can be found in the accompanying paper of Davies et al. 2015, in preparation). The advantage of using the ratios instead of the individual modes of oscillation is that they effectively suppress the contributions from the poorly modelled surface layers \citep[see e.g.,][]{Roxburgh:2005hs,OtiFloranes:2005ii,SilvaAguirre:2011jz}, and are insensitive to the line-of-sight Doppler velocity shift \citep{Davies:2014fe}. This allows direct comparison of theoretical frequencies with the observed quantities without resorting to a {\it surface correction}, such as the power-law term applied to a solar model to reproduce the helioseismic observations \citep[e.g.,][]{Kjeldsen:2008kw}. Moreover, it has been shown that use of the frequency ratios provides the most precise results on asteroseismic ages as compared to fitting individual frequencies or using the scaling relations \citep{SilvaAguirre:2013in,Lebreton:2014gf}.

The ratios and their corresponding covariance matrices have been obtained directly from the MCMC chains following the procedure described in Davies et al.~(2015, in preparation). It is customary to write the ratios $r_{01}(n)$ and $r_{10}(n)$ as one unique set of observables, called $\runo$ \citep[see,][]{SilvaAguirre:2013in}:
\begin{equation}\label{eqn:r010}
r_{010}=\{r_{01}(n),r_{10}(n),r_{01}(n+1),r_{10}(n+1),...\}\,.
\end{equation}

While the frequency ratios provide a reliable proxy of the conditions in the deep stellar interior, their formulation is based on the assumption that the oscillation frequencies behave purely as pressure modes \citep[][and references therein]{Roxburgh:2000js,Roxburgh:2003bb}. Once a star exhausts its central hydrogen supply at the end of the main-sequence phase, its core contracts rapidly resulting in coupling between the propagation cavities of purely acoustic p~modes and gravity g~modes. Consequently, an oscillation mode can take mixed character and behave as a g~mode in the central parts of the star while retaining its p-mode character in the outer layers; this feature is clearly visible in the \'echelle diagram \citep{1983SoPh...82...55G} as an {\it avoided crossing} \citep{Aizenman:1977wh}.

The surface amplitudes of these mixed modes are smaller than those of pure p-modes and they are therefore harder to detect. Two of the 35 targets show clear evidence of modes with mixed character in  their $\ell=1$ frequencies, for which the ratios are not applicable. Their analysis will be presented elsewhere using tools especially designed for dealing with these types of oscillation modes \citep[e.g.,][]{Deheuvels:2011fn,Benomar:2012kv}. A few of the remaining 33 stars show some hint of mixed modes starting to develop or seemingly present in their $\ell=2$ ridges. In these cases only the quantity $\runo$ was matched as described in Section~\ref{ssec_bayes}. Figure~\ref{fig:ech_mix} shows the \'echelle diagram of two KOIs from our sample. The mixed-mode nature in the $\ell=2$ ridge around $\sim$1070~$\mu$Hz is very clear in the case of Kepler-36, while the ridge is completely smooth (and thus p-mode like) in the less evolved star Kepler-409.

\begin{figure}
\begin{center}
\includegraphics[width=\linewidth]{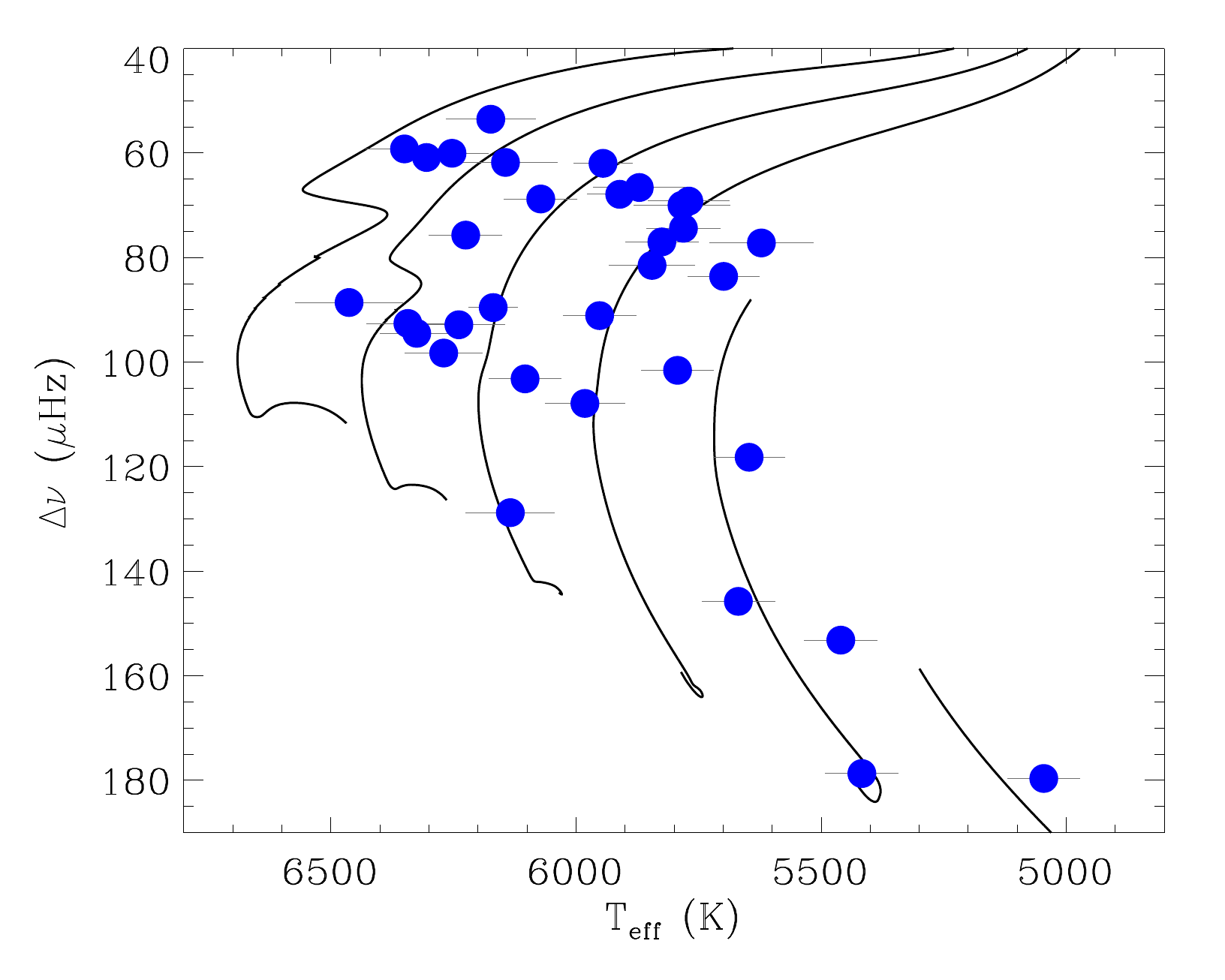}
\caption{Sample distribution in the $\teff$ v/s $\dnu$ diagram, depicted over stellar evolutionary tracks at solar metallicity for masses between 0.8-1.3 M$_\odot$. Observational uncertainties in $\dnu$ are smaller than the symbol size.}
\label{fig:seis_hrd}
\end{center}
\end{figure}

The final sample of stars included in this study is shown in Fig.~\ref{fig:seis_hrd} together with theoretical stellar evolution tracks. Our targets span a wide range of effective temperatures ($5050<\teff<6450$~K), metallicities ($-0.4<\feh<+0.30$~dex), and evolutionary stages that will require the inclusion of different input physics to correctly determine their intrinsic properties.
\section{Stellar models and statistical analysis}\label{sec_par}
The following sections describe the physical processes included in our stellar models, the Bayesian approach, assumptions and observables used to determine the central values and statistical uncertainties, and the adopted set of spectroscopic constraints of our targets. To test the effects in our derived stellar properties of different input physics, as well as properly accounting for systematic uncertainties in our results, we computed several grids of evolutionary tracks changing one physical ingredient at the time.
\subsection{Grids of evolutionary models}\label{ssec_grids}
\begin{table*}
\caption{Description of the input physics considered in each of the grids of models. For all grids, mass separation is 0.01~$\msun$ and metallicity separation is 0.05~dex. When diffusion is included this is done using the prescription of \citet{Thoul:1994iz}. Overshooting is incorporated as an exponential decay of the convective velocities \citep[see,][]{Weiss:2008jy}. Its efficiency has been calibrated with open clusters to $\xi=0.016$ \citep{Magic:2010iz}.}
\label{tab:grids}
\begin{tabular}{cccccccc}
\hline
Name & Mass ($\msun$) &$\feh$ & Solar Abundances & Diffusion &
Overshooting & $\alpha_{\rm MLT}$ & $\Delta Y/\Delta Z$ \\
\hline
\smallskip
GS98sta & 0.70\,;\,1.80 &  +0.50\,;\,-0.65 & \citet{Grevesse:1998cy} & No  & No &  1.791 & 1.4 \\
\smallskip
GS98ove & 1.00\,;\,1.80 &  +0.50\,;\,-0.65 & \citet{Grevesse:1998cy} & No  & Yes &  1.791 & 1.4\\
\smallskip
GS98dif & 0.70\,;\,1.20 &  +0.50\,;\,-0.65 & \citet{Grevesse:1998cy} & Yes  & No &  1.791 & 1.4\\
\smallskip
GS98al$+$ & 0.70\,;\,1.80 &  +0.50\,;\,-0.65 & \citet{Grevesse:1998cy} & No  & No &  2.091 & 1.4\\
\smallskip
GS98al$-$ & 0.70\,;\,1.80 &  +0.50\,;\,-0.65 & \citet{Grevesse:1998cy} & No  & No &  1.491 & 1.4\\
\smallskip
AS09 & 0.70\,;\,1.80 &  +0.50\,;\,-0.65 & \citet{Asplund:2009eu} & No  & No &  1.865 & 1.4\\
\hline
\end{tabular}
\end{table*}
We have computed grids of stellar models using the Garching Stellar Evolution Code \citep[GARSTEC,][]{Weiss:2008jy}, summarised in Table~\ref{tab:grids}. The basic input physics common to all grids includes the 2005 version of the OPAL Equation of State \citep{Rogers:1996iv,Rogers:2002cr}, OPAL opacities for high temperatures \citep{Iglesias:1996dp} and those of \citet{Ferguson:2005gn} for low temperatures, the NACRE compilation of nuclear reaction rates \citep{Angulo:1999kp} including the updated $^{14}\mathrm{N}(p,\gamma)^{15}\mathrm{O}$ reaction from \citet{Formicola:2004dl}, and the mixing-length theory of convection as described in \citet{2012sse..book.....K}.

In our computations the quantity $\feh$ is defined as $\feh=\log(Z_\mathrm{sur}/X_\mathrm{sur})-\log(Z_\mathrm{sur}/X_\mathrm{sur})_\odot$, where $Z_\mathrm{sur}$ and $X_\mathrm{sur}$ correspond to the surface heavy element and hydrogen fractions, respectively. This relation between $\feh$ and $Z/X$ is only valid if the element distribution relative to iron is the same in the observed star as in the Sun. A measurement of alpha abundances $\afe$ is available from high-resolution spectroscopy for one of our targets \citep[Kepler-444, see][]{Campante:2015ei}. Their obtained average value of $\afe=+0.26\pm0.07$ has been transformed to $Z/X$ using the prescription of \citet{Salaris:1993iv}.

The initial compositions of our evolutionary models were determined using a helium-to-metal enrichment law anchored to the Big Bang nucleosynthesis primordial values of $Z_0=0.0$ and $Y_0=0.248$ \citep{Steigman:2010gz}. A linear fit to the initial abundances of a solar calibration results in a relation close to $\Delta Y/\Delta Z=1.4$, which is the one adopted in our grids \citep[see e.g.,][]{Pietrinferni:2004im}. The impact on stellar parameters of variations in the ratio between helium and heavy elements is explored in Section~\ref{ssec_helium}.

The efficiency of convection was determined from a standard solar calibration using microscopic diffusion \citep{Thoul:1994iz} and the \citet{Grevesse:1998cy} solar mixture, resulting in $\alpha_{\rm MLT}=1.791$. Since GARSTEC does not include the effects of radiative levitation or turbulent mixing below the base of the convective envelope, we have restricted the calculations of the grid with diffusion to masses below 1.2~$\msun$ \citep[see][]{1998ApJ...504..559T}. Our diffusion grid comprises masses between 0.7~$\msun$--$1.2~\msun$ in steps of 0.01~$\msun$ and initial compositions of $-0.65<\feh$$<+0.50$ in steps of 0.05~dex. This mass range has been extended in our standard grid (i.e., without diffusion) to 0.7~$\msun$--$1.8~\msun$, also in steps of 0.01~$\msun$. Note that we have kept the value of $\alpha_{\rm MLT}$ constant in these grids because solar models calibrated with and without diffusion show variations in this parameter of $\Delta\alpha_{\rm MLT}\sim0.12$ \citep[see, e.g.,][]{1997A&AS..123..449C}. Our intention is to explore systematic uncertainties by changing one parameter at the time, and as part of this effort we explore larger changes in $\alpha_{\rm MLT}$ (see below).

Furthermore, we constructed a grid including convective overshooting in the diffusive approach implemented in GARSTEC \citep[see Section~3.1.5 in][]{Weiss:2008jy}, using an efficiency of $\xi=0.016$ that has been calibrated to open clusters \citep[e.g.,][]{Magic:2010iz,SilvaAguirre:2011jz} and is roughly equivalent to the $\alpha_\mathrm{ov}=0.2$ pressure scale heights commonly obtained when instantaneous mixing in the overshoot region is used. Additional grids of models were constructed changing the value of the mixing-length parameter and the reference set of solar abundances (called GS98al$+$, GS98al$-$, and AS09 in Table~\ref{tab:grids}, see also Section~\ref{ssec_phys} below). The input physics of the grids is detailed in Table~\ref{tab:grids}, and we will use the naming convention given there for the remainder of this paper.

For hundreds of models along each evolutionary track, we calculated theoretical oscillations frequencies using the Aarhus Adiabatic Oscillation Package \citep[ADIPLS,][]{ChristensenDalsgaard:2008kr}. Each grid comprises over half a million models fully covering the spectroscopic and asteroseismic parameter space of our targets. The final stellar properties for our targets have been obtained with either the GS98dif or the GS98ove grids, for reasons thoroughly discussed in Sections~\ref{ssec_dif} and~\ref{ssec_ove} below.
\subsection{Bayesian approach}\label{ssec_bayes}
The stellar properties of each star are determined using an adapted version of the Bayesian approach described in \citet{Serenelli:2013fz} that we have named the BAyesian STellar Algorithm (BASTA). If $\mathbf{v}$ is a set of model stellar parameters (e.g. mass, composition, age), and $\mathcal{O}$ the observed data, then according to Bayes' rule the probability density function (PDF) of $\mathbf{v}$ given $\mathcal{O}$ is
\begin{equation}\label{eqn:pdf} 
p\left(\mathbf{v} | \mathcal{O}\right) \propto p\left(\mathbf{v}\right) \mathcal{L}\left( \mathcal{O}|\mathbf{v} \right).
\end{equation} 
Here, $\mathcal{L}\left(\mathcal{O}|\mathbf{v}\right)$ is the likelihood of $\mathcal{O}$ given $\mathbf{v}$ and $p\left(\mathbf{v}\right)$ is the prior PDF of $\mathbf{v}$ representing any prior knowledge on these quantities.

To determine $p\left(\mathbf{v}\right)$ we considered the following. The studied sample is comprised of stars presenting transit-like signals in their light curves, as well as stochastically excited oscillations. Since the selection function of pulsating exoplanet candidate host star observed by \emph{Kepler} is complex, we cannot determine a prior probability for stellar properties based on these characteristics. Therefore, we assume a flat prior in [Fe/H] and age including only a strict cut on the latter at 15~Gyr (to properly construct the age PDF of old stars), and we use a standard Salpeter Initial Mass Function.

We compute the likelihood of the observed values $\mathcal{O}$ given a set of model parameters $\mathbf{v}$ assuming Gaussian distributed errors. BASTA is flexible in its input, and the observables included can be the spectroscopic parameters ($\teff$, $\feh$, $\logg$), the individual oscillation frequencies, the global average asteroseismic parameters $\langle\dnu\rangle$ and $\num$, the frequency ratios $\rdos$ and $\runo$ defined in Equations~\ref{eqn:r02}~and~\ref{eqn:r010}, or any combination of the aforementioned quantities. The construction of the ratios introduces correlations as a function of frequency which are taken into account in the likelihood calculation of each model:
\begin{eqnarray}
\mathcal{L}\left( \mathcal{O}|\mathbf{v} \right) & = & \frac{1}{(2\pi)^{1/2}\sqrt{\bf \left |C \right |}} \exp{\left(-\chi^2/2\right)}\,,\label{eqn:like}\\
\chi^2 & = & \left(\vec{o}_{\mathrm{obs}} - \vec{o}_{\rm model}\right)^{T} \, {\bf C}^{-1} \, \left(\vec{o}_{\mathrm{obs}} - \vec{o}_{\rm model}\right)\,,\label{eqn:chi2}
\end{eqnarray}
where {\bf C} is the covariance matrix of the observed values $\vec{o}_{\mathrm{obs}}$, and $\vec{o}_{\mathrm{model}}$ are the same quantities determined from the model. For computational efficiency we preselect models for determining the likelihood within 500~K and 0.5~dex of the spectroscopic $\teff$ and $\feh$ values, and having an average large frequency separation $\langle\dnu\rangle$ within 15\% of the observed one (approximately 3-$\sigma$ from the median uncertainty for the majority of the asteroseismic {\it Kepler} solar-type sample, see \citet{Chaplin:2014jf}). The value of $\langle\dnu\rangle$ in the models is obtained using theoretical frequencies of oscillation following the prescription of \citet{White:2011fw} (see section~\ref{sec_scal} below).

Our final values for stellar properties of all targets were determined with BASTA and GARSTEC grids using $\teff$, $\feh$, and the frequency ratios $\runo$ and $\rdos$ as the observables included in the likelihood calculation (except in a few cases where only $\runo$ was fitted because $\ell=2$ mixed modes prevented us from using $\rdos$, see section~\ref{sec_targ}). For comparison we have also determined all stellar properties using the individual oscillation frequencies (see Section~\ref{ssec_meth}) and the global average parameters $\langle\dnu\rangle$ and $\num$ (see Section~\ref{sec_scal}) as the asteroseismic data to be fitted.

Recently, \citet{Roxburgh:2013ev} proposed that fitting the separation ratios $\runo$ and $\rdos$ as a function of radial order could lead to biases in the determined stellar structure, and recommended comparing observations with models by interpolating the theoretical ratios at the observed central frequencies. We have implemented this fit in BASTA and found no significant difference with the results obtained when comparing at a fixed radial order. Similar conclusions were given by \citet{Lebreton:2014gf} (see their Appendix A.1).

Once the probability density function has been computed using the definitions above, the marginalised posterior PDF of any stellar quantity $x$ is obtained as
\begin{equation}\label{eqn:poster} 
p\left(x|\mathcal{O}\right) = \int{\delta(x(\mathbf{v})-x) p\left(\mathbf{v} | \mathcal{O}\right) }w_v d^3v, 
\end{equation}
where $w_v$ are the appropriate weights used to account for the volume of parameter space occupied by a stellar track point characterised by $\mathbf{v}$, i.e. half the distance to its neighbouring points in mass, metallicity, and age. The final values reported for each stellar quantity are obtained from the median of the posterior PDF and the 16 and 84 percentiles.
\subsection{Spectroscopic input parameters}\label{ssec_spe}
\begin{table*}
\caption{Spectroscopic determinations of selected KOIs. The $\logg$ values were adopted from asteroseismic determinations of \citet{Huber:2013jb}. Uncertainties quoted in $\teff$ and $\feh$ correspond to the internal precision of the method.}
\label{tab:spectra}
\begin{tabular}{ccccccc}
Name & Spectrum & S/N & $\teff$ (K) & $\logg$ (dex) & $\feh$ (dex) & $\xi_{\rm turb}$~(km/s) \\ 
\hline
\smallskip
KOI-5  &  McD2.7m &    70 &  5945$\pm$60 &  4.05  & +0.17$\pm$0.05  & 1.3\\
\smallskip
Kepler-21 &  TRES  &  150 &  6305$\pm$50    &  4.02  & -0.03$\pm$0.05   & 1.6\\
\hline
\end{tabular}
\end{table*}
We initially adopted the spectroscopic values reported in \citet{Huber:2013jb} as input parameters in the Bayesian procedure described above. Since this approach allows determination of the posterior PDF from models of any stellar quantity, comparison showed that in two stars the best asteroseismic solutions favoured atmospheric parameters not compatible with the spectroscopic ones. To explore the reasons for this discrepancy, we have made a new determination of $\teff$ and $\feh$ from high-resolution spectra on these single targets as available on the website of the Kepler Community Follow-up Observing Program (CFOP\footnote{https://cfop.ipac.caltech.edu/home/}).

The method takes advantage of the accurate values of the stellar surface gravity, $\logg$, determined from seismic data, which allows a determination of $\teff$ from the requirement that the FeI and FeII lines should give the same Fe abundance. Given that $\feh$ derived from FeI lines {\em increases} by $\sim$0.055~dex for solar-type stars when $\teff$ is increased by 100~K, and that $\feh$ based on FeII lines {\em decreases} by $\sim$0.03~dex for the same change in $\teff$, the method leads to precise values of $\teff$~($\pm$70~K) if the two $\feh$ values are determined with errors less than $\pm$0.03 dex.

This is possible when high-resolution spectra having S/N~$>$~40 are analysed differentially with respect to the Sun. Based on the solar spectrum we have selected 39 FeI and 14 FeII unblended lines in the wavelength region 480$-$650~nm. Their equivalent widths (EWs), which range from 30 to 80 m{\AA} in the stellar spectra, were measured by Gaussian fitting to the line profiles and analysed with MARCS model atmospheres \citep{Gustafsson:2008df} to provide LTE abundances as described in \citet{Nissen:2014cj}. The Fe abundances from FeI lines are subject to small non-LTE effects, which were taken into account using the calculations of \citet{Lind:2012gc}. These corrections decrease the derived $\teff$ by 15 to 30~K.

The abundances are based on medium to strong spectral lines that are sensitive to the microturbulence velocity $\xi_{\rm turb}$. This parameter is not well constrained from the set of lines used, and we have therefore adopted a relation $\xi_{\rm turb}$~(km/s)$= 1.0 + 0.00063 (\teff-5777)  - 0.54 (\logg - 4.44)$ derived for solar-type stars with high S/N HARPS spectra \citep{Nissen:2014cj}. The derived $\teff$ is not very sensitive to $\xi_{\rm turb}$, because the two sets of FeI and FeII lines have about the same average EW. For an estimated error of 0.2 km/s in  $\xi_{\rm turb}$, the error in $\teff$ is 25~K, whereas the corresponding error in the derived $\feh$ is 0.05~dex. Furthermore, we note that an error of 0.05 dex in $\logg$ induces an error of 20 K in $\teff$, and has a negligible effect ($<$0.01~dex) on $\feh$.
 
The results obtained are given in Table~\ref{tab:spectra}. The listed errors correspond to the uncertainties in $\feh$ estimated from the line-to-line scatter of the derived abundances. The small errors arising from the uncertainties in $\logg$ and $\xi_{\rm turb}$ were added in quadrature. For the two stars in Table~\ref{tab:spectra}, $\teff$ and $\feh$ deviate significantly from the values determined in previous studies, i.e. KOI-5 \citep[$\teff=5753\pm75$~K, $\feh=0.05\pm0.1$~dex,][]{Huber:2013jb}, and Kepler-21 \citep[$\teff=6131\pm44$~K, $\feh=-0.15\pm0.06$~dex,][]{Howell:2012jv}. The final stellar properties in this paper were determined using these updated results. We emphasise that, for completeness, we analysed another five stars from the \citet{Huber:2013jb} sample and found good agreement: average differences and standard deviations for these targets (new values$-$\citet{Huber:2013jb}) are $\Delta\teff = -13\pm70$~K and $\Delta\feh=-0.06\pm0.06$~dex.
\section{Results}\label{sec_res}
To determine the final stellar properties and uncertainties of our sample, we have devised the following procedure. The central values and statistical uncertainties were obtained using the BASTA and a choice of GARSTEC grids including different input physics as described in Section~\ref{ssec_cen}. These are the recommended stellar properties of the sample and are listed in Table~\ref{tab:stellar}. We then explored the impact of changing the reference solar abundances and mixing-length parameter (section~\ref{ssec_phys}) to estimate a systematic uncertainty from the input physics. We then compared our results with those obtained from three different asteroseismic pipelines, providing a systematic uncertainty arising from the use of different evolution codes and fitting algorithms (section~\ref{ssec_meth}). Finally, we estimated a systematic uncertainty arising from our choice of $\Delta Y/\Delta Z=1.4$ (section~\ref{ssec_helium}), and discuss the possibility of breaking the degeneracy between helium and mass using \emph{Gaia} results.
\subsection{Adopted physics for stellar properties and statistical uncertainties}\label{ssec_cen}
\begin{figure}
\begin{center}
\includegraphics[width=\linewidth]{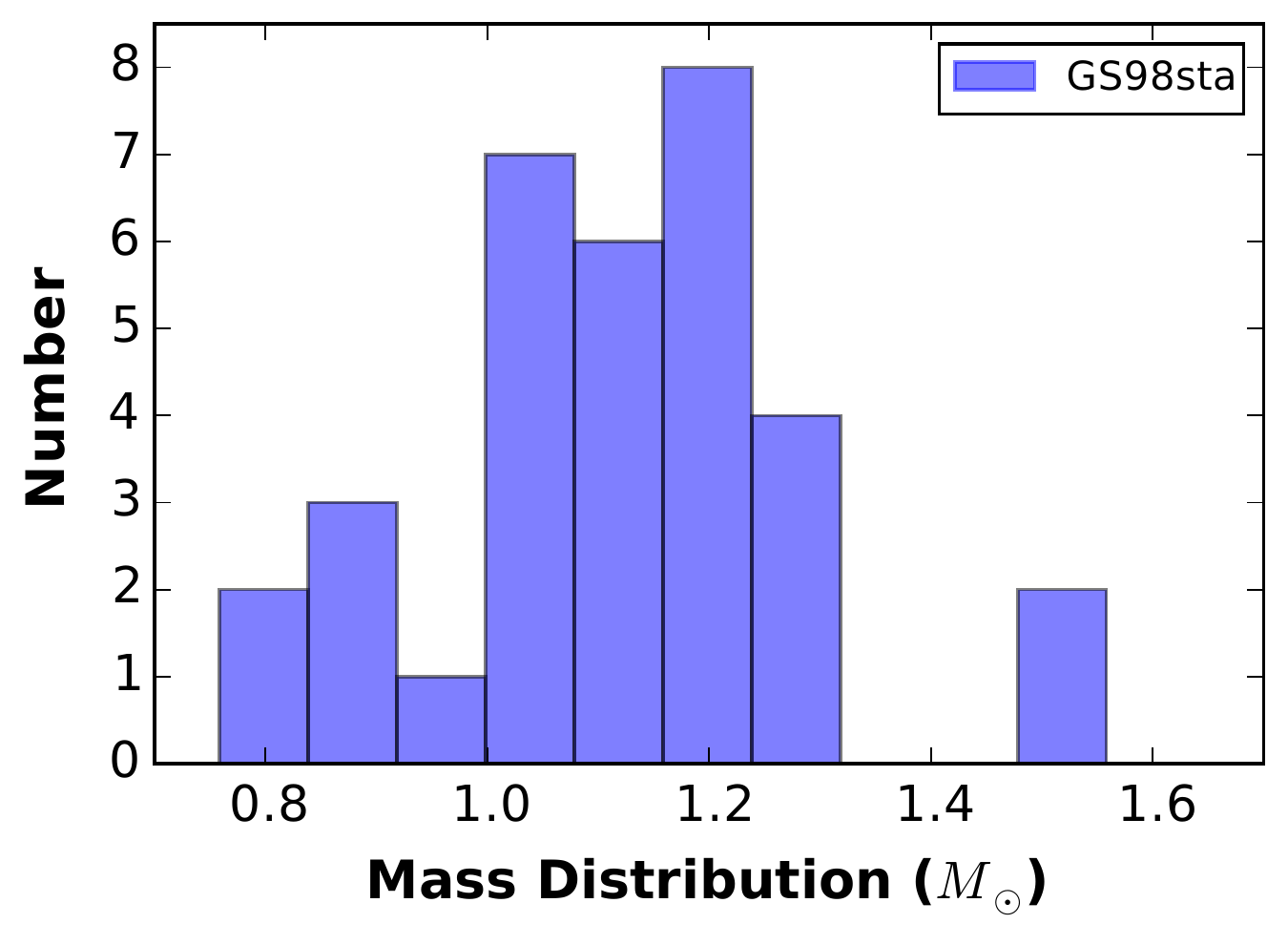}
\caption{Mass distribution of the sample obtained using the standard grid (GS98sta).}
\label{fig:mass_dist}
\end{center}
\end{figure}

The BASTA described in Section~\ref{ssec_bayes} was initially applied to the sample using the standard grid (GS98sta, see Table~\ref{tab:grids}). Figure~\ref{fig:mass_dist} shows the obtained distribution of masses for the targets, encompassing a range between $\sim$0.75$-$1.6~$\msun$. There are several physical processes that can considerably affect the main-sequence evolution of stars, but whose efficiency and range of applicability is not fully constrained by theory or independent observations. We explore in the following sections the cases of microscopic diffusion and overshooting.
\subsubsection{Microscopic diffusion}\label{ssec_dif}
In the low-mass regime, neglecting the effects of microscopic diffusion can have significant impact on the determined properties. If stars are indeed subject to element settling throughout their main-sequence lifetimes, the derived ages can be particularly biased (e.g., Valle et al. 2014b). Evidence for the occurrence of microscopic diffusion is clear for the Sun \citep[e.g.,][]{1993ApJ...411..394G,ChristensenDalsgaard:1993ck} and has been growing in recent years for star clusters \citep[see e.g.,][]{2007ApJ...671..402K,Gruyters:2013jp,Onehag:2014jd} revealing the importance of this process in low-mass stars.

\begin{figure}
\begin{center}
\includegraphics[width=84mm]{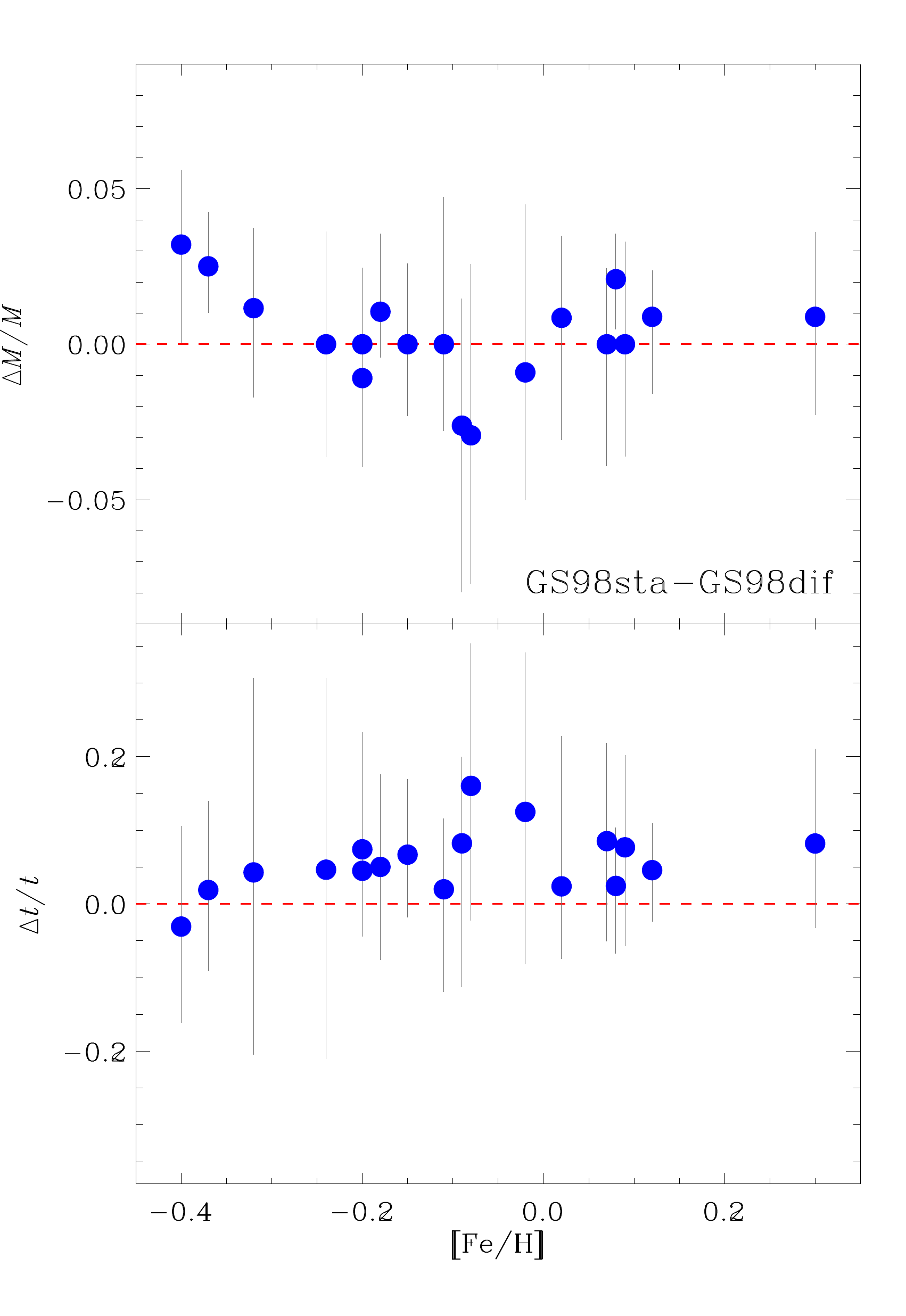}
\caption{Fractional differences in mass and age obtained from the standard and diffusion grids of models as a function of spectroscopic metallicity, in the sense (GS98sta-GS98dif) {\it Top}: mass comparison. {\it Bottom}: age comparison. See text for details.}
\label{fig:norm_diff}
\end{center}
\end{figure}

Using the grid of models that includes the effects of diffusion (GS98dif) we have redetermined the stellar properties of all stars with masses predicted from the standard grid to be below $\sim$1.2~$\msun$. The top panel of Fig.~\ref{fig:norm_diff} shows the resulting fractional mass differences, with the two determinations typically agreeing within $\sim$2.5\%. There is no visible trend as a function of metallicity although the largest differences are found for the most metal-poor stars, as expected \citep[see e.g.,][]{Valle:2014fs}. However, the ages predicted by the standard grid are systematically older than those based on diffusion (bottom panel in Fig.~\ref{fig:norm_diff}). Since the masses determined from both grids agree on average, this behaviour is not what would be expected from a pure mass effect and is a consequence of several physical effects combined that balance each other.

Stellar models that include microscopic diffusion sink helium and heavy elements while diffusing hydrogen towards the surface during their main-sequence lifetime. This process effectively reduces the amount of fuel available in the core producing a faster main-sequence evolution and therefore decreasing the age a model has at a given mean density. Interestingly the effects of diffusion on age counterbalance those expected from a pure mass effect, since even if the mass predicted by the non-diffusive model is higher than the diffusive one its age remains older. Thus, despite not producing biases in the masses determined, the use of diffusion does affect the ages obtained from asteroseismic fitting and should be taken into account in the relevant regime. Similar results were found by e.g.,~\citet{Miglio:2005is,2014arXiv1412.5895V}, where models with diffusion predict younger ages than those not including this effect.

\begin{figure*}
\includegraphics[width=84mm]{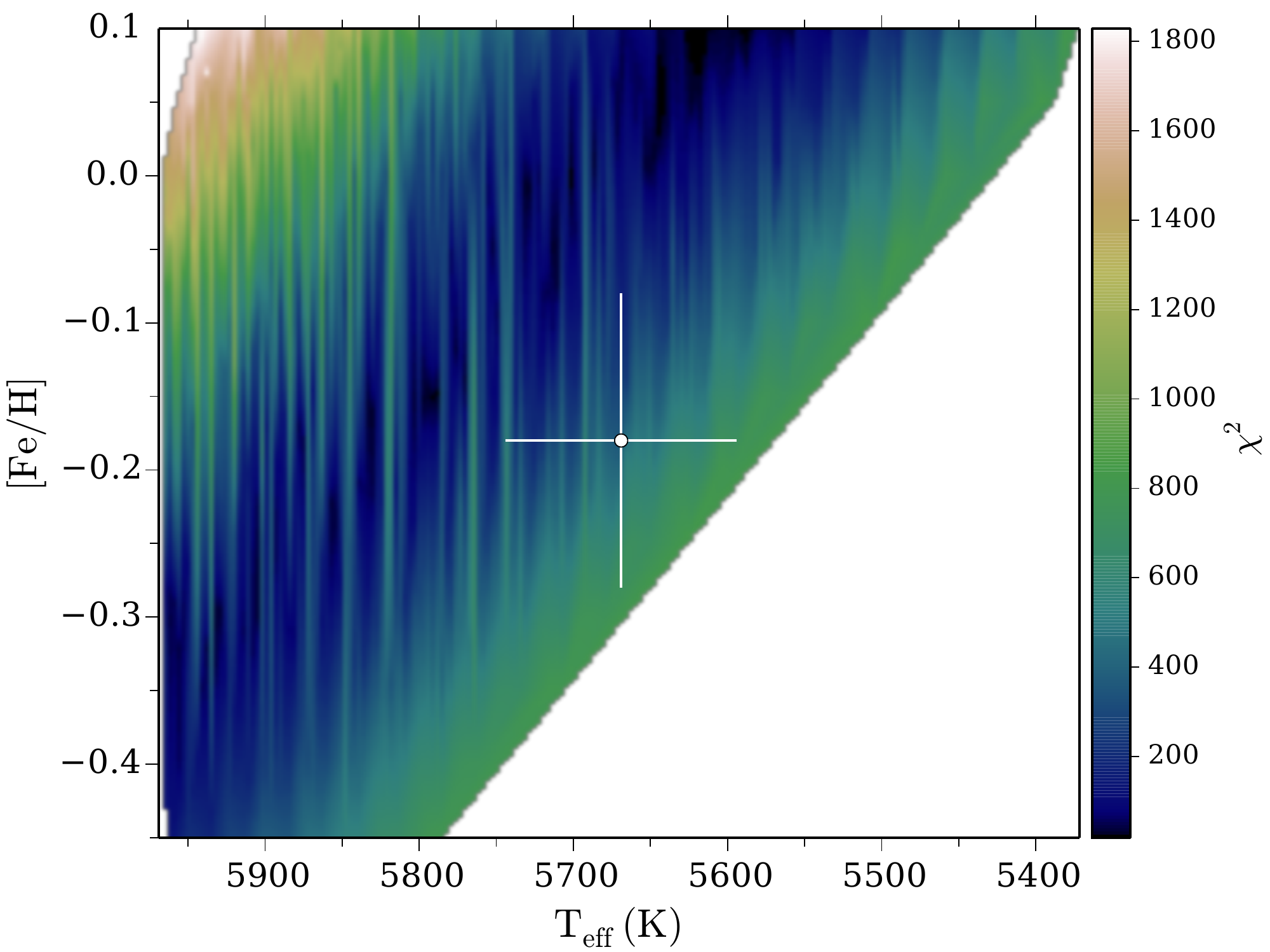}
\includegraphics[width=84mm]{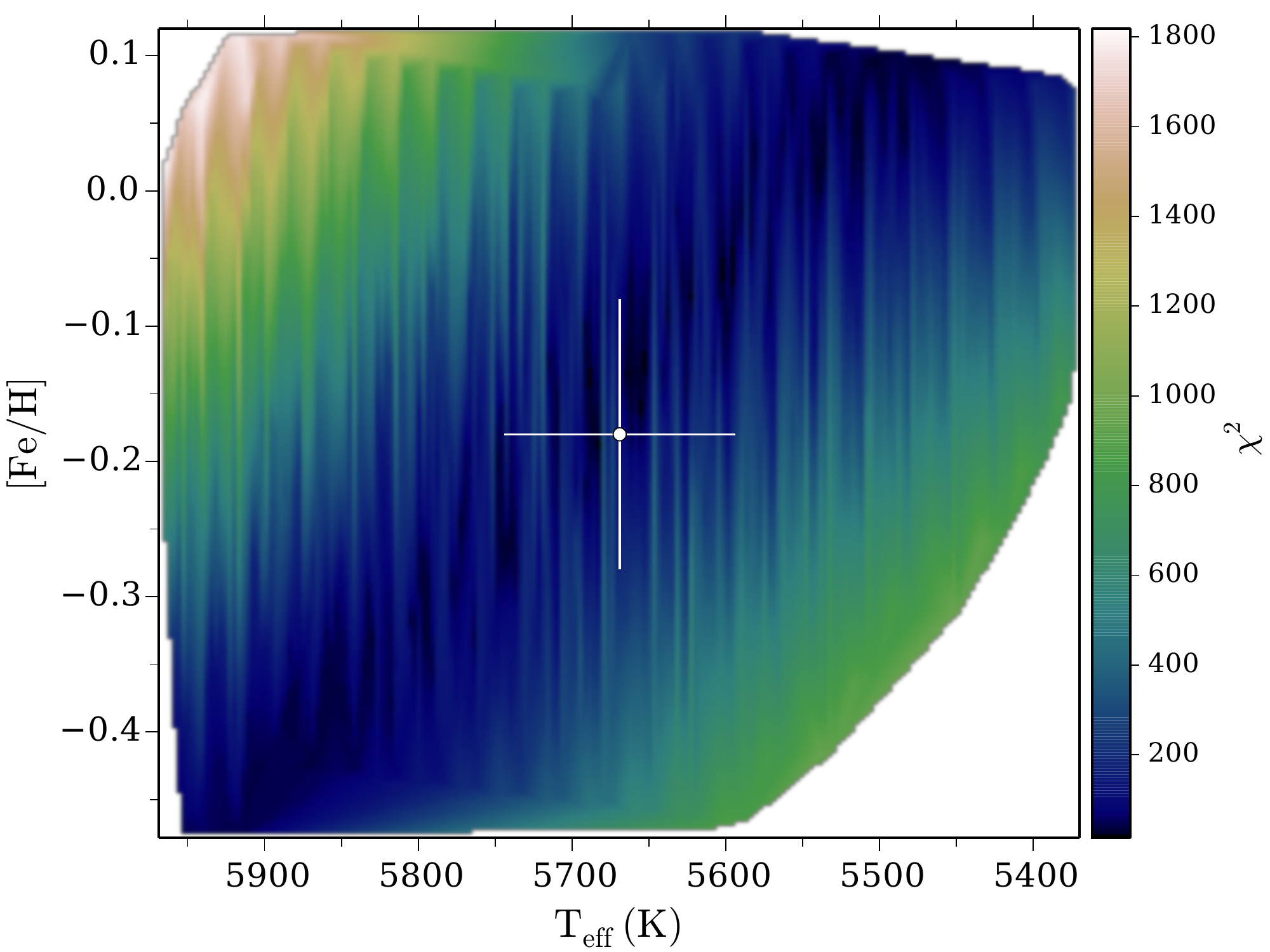}
\caption{Values of $\chi^2$ determined using the frequency ratios and spectroscopic constraints for Kepler-93. Solid lines depict the $1\,\sigma$ uncertainties in the spectroscopic atmospheric parameters. {\it Left}: standard grid. {\it Right}: diffusion grid.}
\label{fig:koi69_diff}
\end{figure*}

Compared to canonical models, the transport of chemical species by microscopic diffusion produces different composition profiles and therefore different opacity values through the stellar interior. As a consequence, evolutionary tracks computed with the same initial chemical mixture including or not these effects will also show a different effective temperature evolution during the main-sequence phase. An example of the effects played by microscopic diffusion is shown in Fig.~\ref{fig:koi69_diff}. These results are obtained for Kepler-93 using the frequency ratios $\runo$, $\rdos$, and spectroscopic parameters as the observational data to be reproduced. The plots depict the $\chi^2$ surface from each grid of models in the $\teff$~versus~$\feh$ plane as well as the observed spectroscopic properties and uncertainties. Due to the number of observables and size of the fractional uncertainties, the fit is dominated by the asteroseismic data and the grid neglecting diffusion favours a parameter space that is not fully compatible with the observed $\teff$ and $\feh$ of the target. In particular, GS98sta results predict $\teff$ values too high compared with the spectroscopic ones. For this reason, the grid with diffusion has been used for determining the final central values and statistical uncertainties of targets with masses below $\sim$1.2~$\msun$.

One must be cautious when determining uncertainties using the diffusion grid because it suffers from an unavoidable edge effect for masses above $\sim$1.15~$\msun$, where the posterior PDF are artificially cut due to the chosen upper mass limit when constructing the grid (1.20~$\msun$). To ensure a smooth transition in the obtained stellar properties when switching from GS98dif to another grid, we compared the results predicted with GS98sta and GS98ove to those of GS98dif in six targets at the edge of this mass range (KOI-5, Kepler-65, Kepler-25, Kepler-126, KOI-268, and Kepler-129, see Table~\ref{tab:stellar}). All stars where the GS98dif grid returned central mass values above 1.15~$\msun$ clearly showed the edge effect in the uncertainties, but the central values in the stellar properties are fully compatible with those obtained from the other two grids (note that the effects of overshooting are not expected to be significant in the 1.15-1.2~$\msun$ range, see Section~\ref{ssec_ove} below). The reason is that for these masses, the evolutionary timescale is short enough for the effects of diffusion to be minor. Thus, in our recommended values presented in Table~\ref{tab:stellar} we restrict ourselves to the diffusion grid in all stars with central mass values below 1.15~$\msun$.
\subsubsection{Convective overshooting}\label{ssec_ove}
In stars with masses higher than $\sim$1.1~$\msun$ the size of the convective core severely affects the age at which central hydrogen exhaustion (the terminal age main-sequence, TAMS) is reached. The extent of the central mixed region is controlled by the convective overshoot efficiency, a free parameter in stellar evolution calculations. There is a wealth of evidence in the literature that convective core sizes are probably larger than those predicted by models applying the pure Schwarzschild criterion for convective boundary determination \citep[e.g.,][just to name a few]{Maeder:1974wc,Maeder:1974tm,Maeder:1991to,Chiosi:1992gp,Vandenberg:2007cq}, and recent analyses of {\it Kepler} stars have further supported this scenario \citep{SilvaAguirre:2013in}. Thus, we have applied our Bayesian method using the overshooting grid (GS98ove) for all targets and compared the results to those from the standard set of models.

\begin{figure}
\begin{center}
\includegraphics[width=84mm]{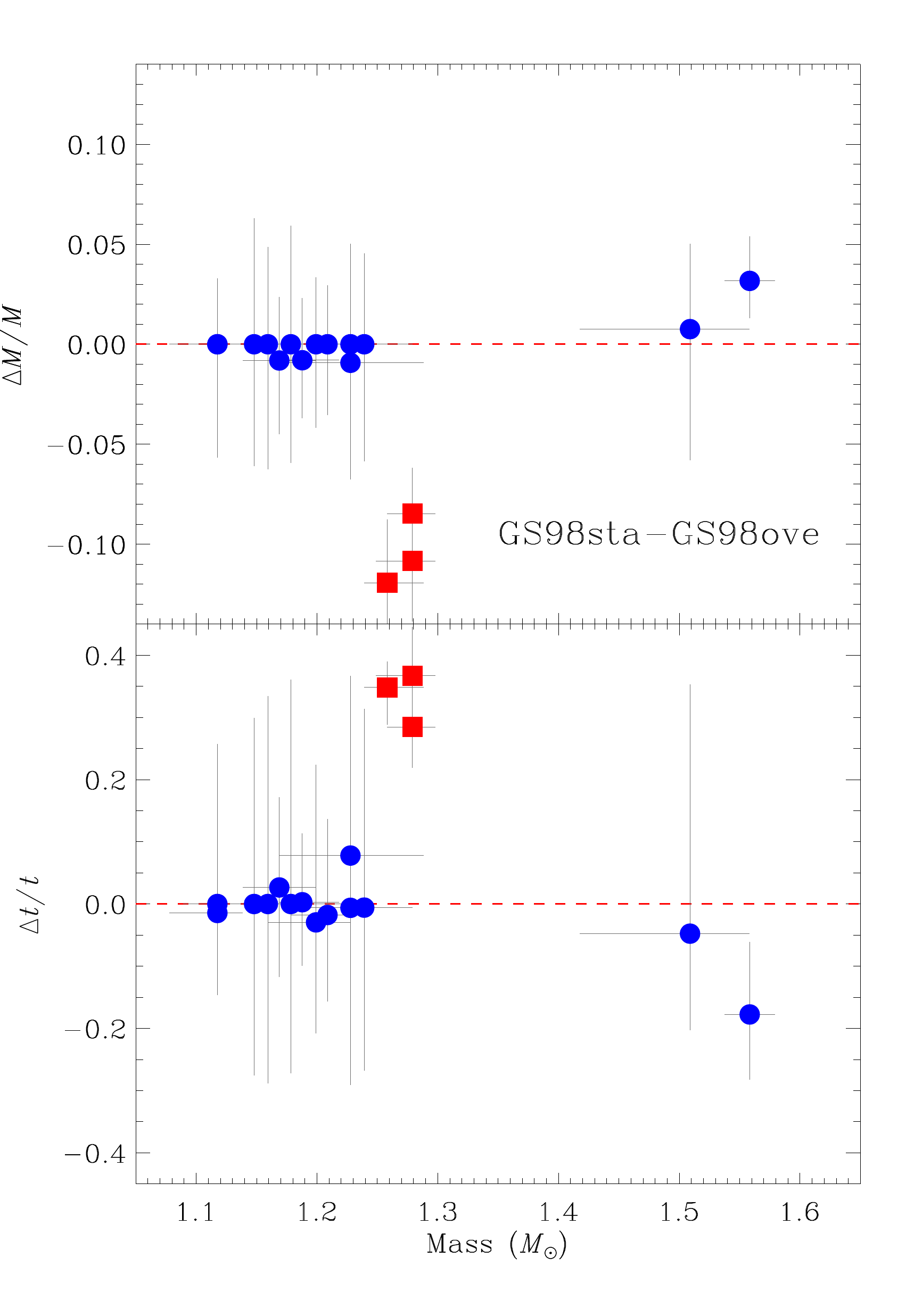}
\caption{Fractional differences in mass and age obtained with the standard and overshooting grids as a function of stellar mass, in the sense (GS98sta-GS98ove) {\it Top}: mass comparison. {\it Bottom}: age comparison. Red squares shows the positions of outliers Kepler-145, KOI-974, and Kepler-21. See text for details.}
\label{fig:normal_ove}
\end{center}
\end{figure}

Figure~\ref{fig:normal_ove} shows the fractional difference in mass and age obtained from both grids for stars with masses above 1.1~$\msun$. In three targets (Kepler-145, KOI-974 and Kepler-21) the mass and age deviations can be attributed to the quality of the fits: the PDFs of these particular stars show a clearly bimodal distribution (see below). A similar situation is observed in the target at $\sim$1.5~$\msun$,  where the large mass and age uncertainty in the GS98sta results is a consequence of its Bayesian results encompassing the secondary peak in the distribution.

\begin{figure}
\begin{center}
\includegraphics[width=84mm]{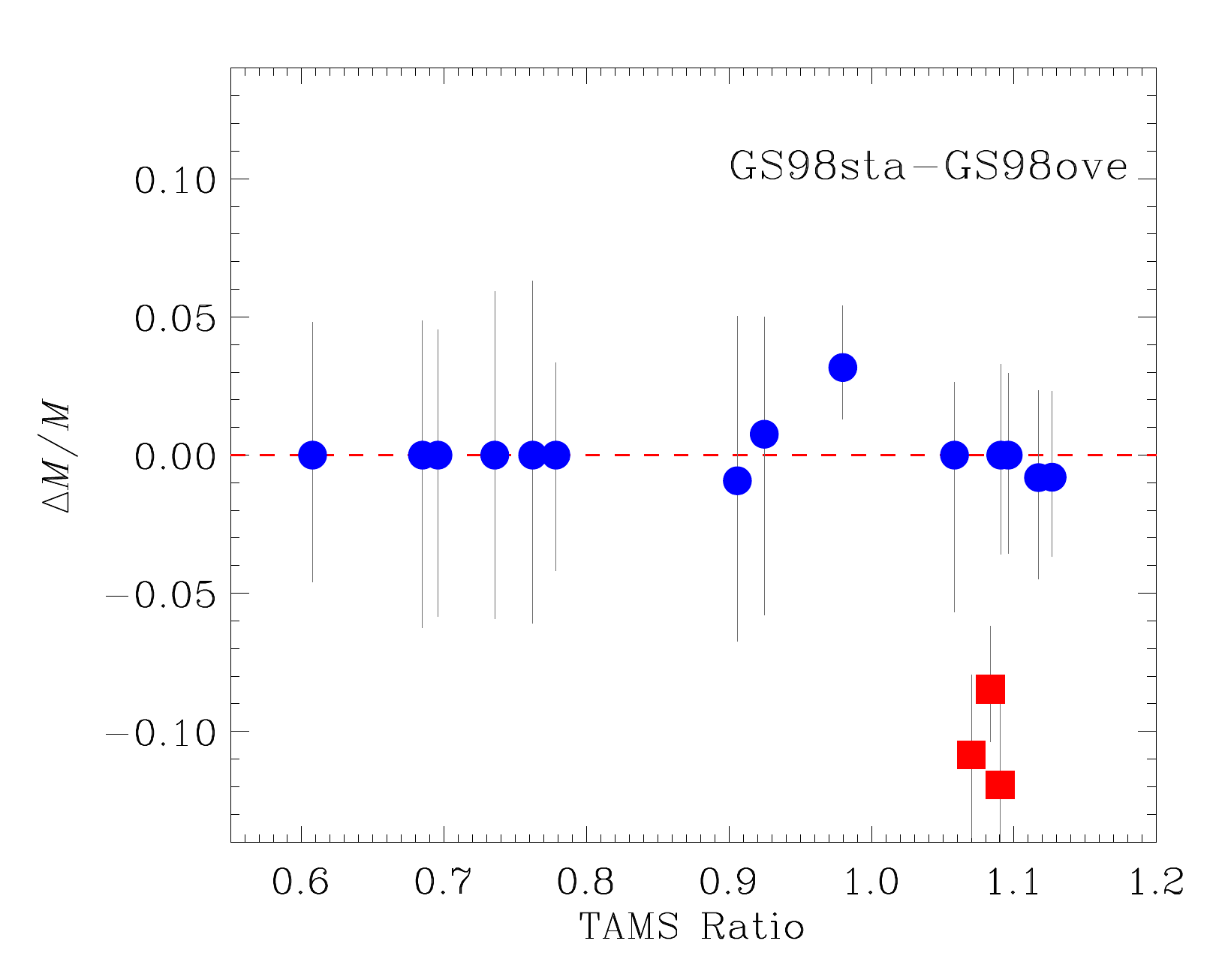}
\caption{Fractional mass difference between the standard and overshooting grids as a function of the Terminal Age Main Sequence (TAMS) ratio for the GS98sta grid. Red squares shows the positions of outliers Kepler-145, KOI-974, and Kepler-21. The TAMS ratio is defined as the age of the best fit model in the GS98sta grid divided by the age at the end of the core hydrogen-burning phase. See text for details.}\label{fig:normal_ove_tams}
\end{center}
\end{figure}

Besides the obvious outliers (red squares in Fig.~\ref{fig:normal_ove}), differences in the results are noticeable in five targets (HAT-P7, KOI-5, Kepler-50, Kepler-129, and KOI-288). Figure~\ref{fig:normal_ove_tams} shows the fractional mass difference as a function of the TAMS ratio, defined as the ratio between the age of each target's best fit model and the age that model has when exhausting its central hydrogen content. The differences in mass appear at TAMS ratio values of $\sim$0.9 when convective cores are well developed and strongly influence the frequency ratios \citep{SilvaAguirre:2011jz}. Moreover, once the stars reach the turn-off at TAMS ratio of $\sim$1.0 some results are fully compatible while others still show small differences. This might be an indication of asteroseismic data keeping record of past mixing processes in stars once these have evolved slightly beyond the main-sequence phase \citep[see][where similar results were found]{Deheuvels:2011fn}. We note that for our targets with TAMS ratio larger than 1, their best-fit models show evidence of mixed modes  with no obvious counterparts in the observations, probably due to their low amplitudes.

\begin{figure}
\begin{center}
\includegraphics[width=84mm]{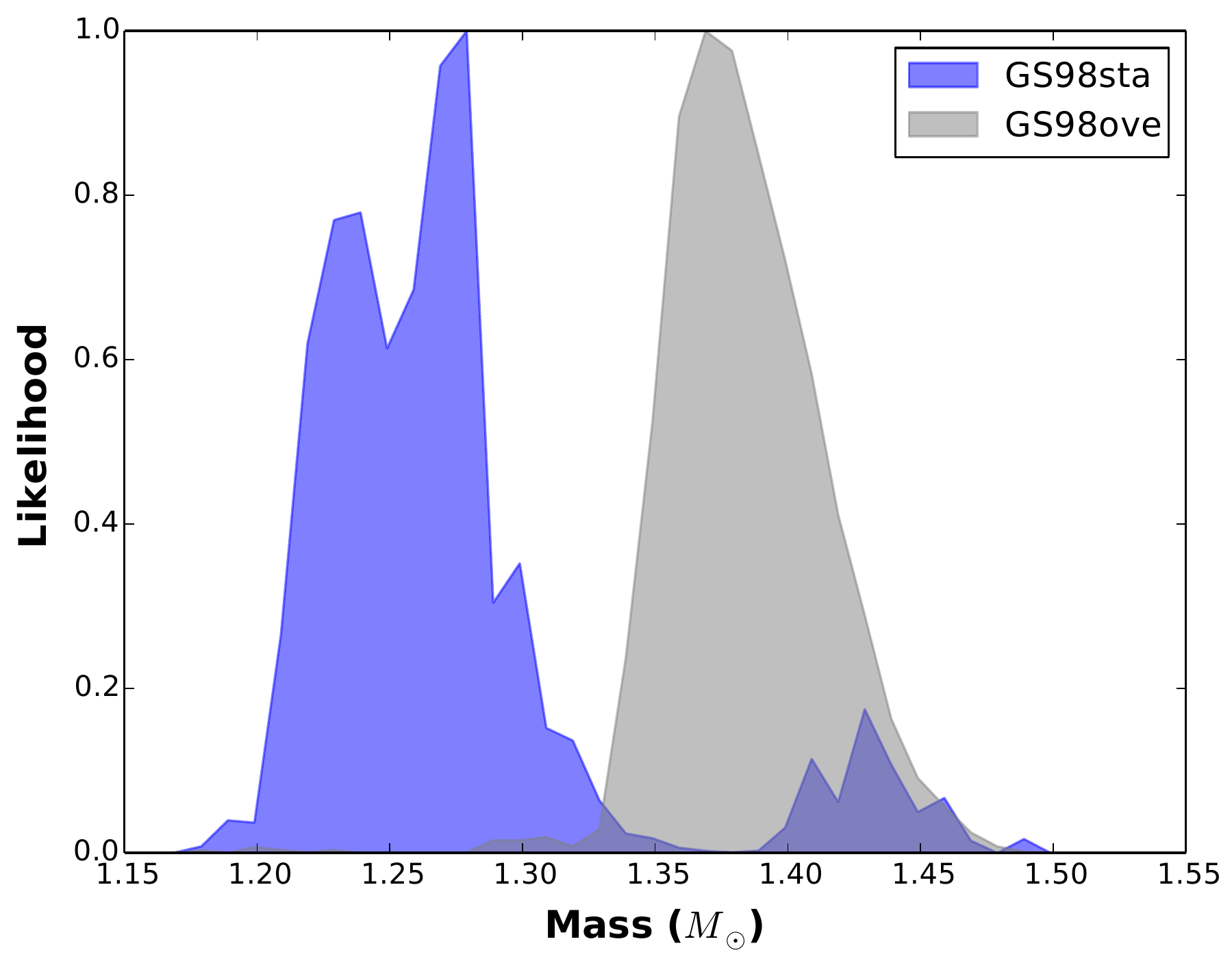}
\caption{Mass distribution obtained with BASTA for KOI-974 using the standard and overshooting grids.}
\label{fig:bimodal}
\end{center}
\end{figure}

The three outliers present in Figs.~\ref{fig:normal_ove} and~\ref{fig:normal_ove_tams} (Kepler-145, KOI-974, and Kepler-21) stem from bimodal probability distribution functions obtained from the GS98sta grid, which are not observed in the results when the GS98ove models are used instead. Figure~\ref{fig:bimodal} shows the example of KOI-974, where the GS98sta grid favours a star with mass $\sim$1.27~$\msun$ that has evolved beyond the main sequence (TAMS$\sim$1.1, see Fig.~\ref{fig:normal_ove_tams}). However, there is evidence of a secondary solution at higher masses, close to $\sim$1.43~$\msun$. When using the overshooting grid instead the PDF shows a unique peak at $\sim1.39$~$\msun$ compatible with the high-mass solution seen in the GS98sta grid. A similar behaviour is observed in the PDFs for Kepler-145 and Kepler-21, where the GS98ove results favour masses consistent with the higher-mass secondary peak in the GS98sta probability distribution function. We note that the bimodal nature of the solution for Kepler-21 had already been pointed out by \citet{Howell:2012jv}.

For the three outliers the inclusion of overshooting produces a better fit to the frequency combinations and argues against the lower-mass peak in the PDF. Since the GS98sta and GS98ove grids cover the same parameter space in terms of initial masses and compositions, we also compared the average likelihood of the grids for these targets and always found better agreement with the overshooting models. In fact, the GS98ove results in all three cases favours stars close to the end of the main-sequence phase (TAMS $\sim$0.92), where convective cores are well developed and masses determined with the standard grid are slightly higher than with overshooting (see Fig.~\ref{fig:normal_ove_tams}). Thus, the final parameters of these three KOIs, given in Table~\ref{tab:stellar}, are those determined from the GS98ove. This solution for Kepler-21 also results in a luminosity and distance compatible with that determined from parallaxes, as described in section~\ref{ssec_helium} below. The actual mass value at which extending the convective core size by means of overshooting has an impact in the obtained properties is expected to be between $1.2-1.3\msun$, but the quality of the data and mass coverage of the current sample are not adequate to thoroughly test this transition. Figures~\ref{fig:normal_ove} and~\ref{fig:normal_ove_tams} show that this reasonable expectation is consistent with the data.
\begin{figure}
\begin{center}
\includegraphics[width=84mm]{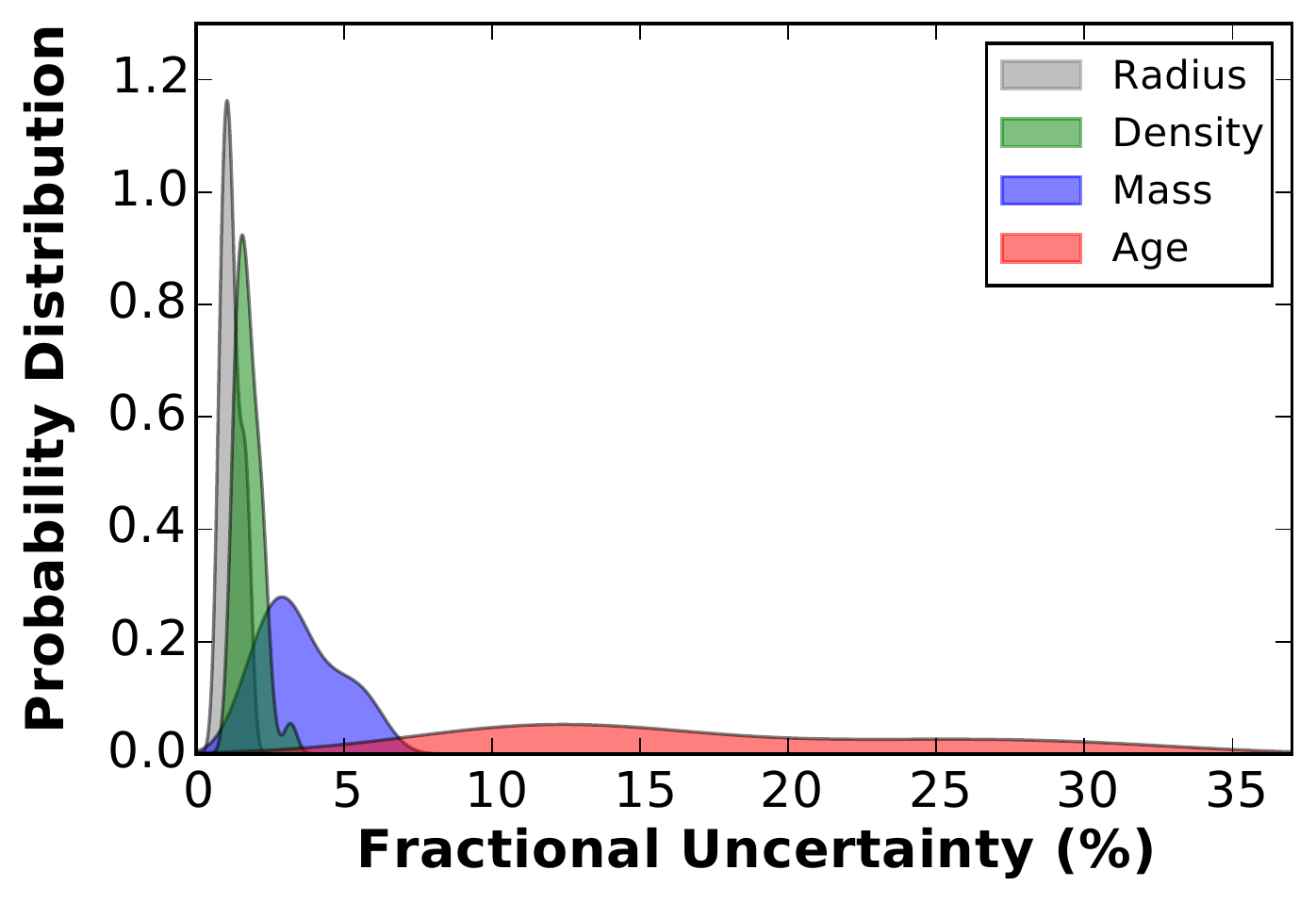}
\caption{Fractional uncertainties added in quadrature in the final stellar parameters determined with our BASTA. See text for details.}\label{fig:final_uncert}
\end{center}
\end{figure}

Based of the analysis and comparisons on the input physics discussed in Sections~\ref{ssec_dif} and~\ref{ssec_ove}, we choose the GS98dif grid to determine the final stellar properties and statistical uncertainties for stars of masses below $\sim$1.15~$\msun$, and the GS98ove grid for the remainder of the sample. These values are given in Table~\ref{tab:stellar}. In terms of internal precision, the median fractional uncertainties returned by the BASTA are 1.2\% (radius), 1.7\% (density), 3.3\% (mass), and 14\% (age). We note that the seemingly large fractional uncertainty on the density compared with asteroseismic determinations made by other methods arises from the use of frequency ratios instead of individual oscillation frequencies. The distributions for our sample shown in Fig.~\ref{fig:final_uncert} correspond to the statistical uncertainties only. To check how robust this error budget is we explore in the next sections the systematic contribution to the uncertainties from the input physics and fitting algorithms.
\subsection{Systematic uncertainties from input physics}\label{ssec_phys}
Our central recommended values and statistical uncertainties derived in the previous sections were based on fixed sets of input physics chosen according to our current best knowledge of stellar evolution (i.e., including microscopic diffusion and core overshooting in the relevant mass regimes). In this section we explore the impact on our results from changing two quantities whose values are not yet fully constrained by observations or theory, namely the solar abundances and the mixing-length parameter.

The photospheric composition of the Sun has been revised in recent years inspired by the predictions of 3D model atmospheres \citep[e.g.,][]{Asplund:2005uv, Asplund:2009eu,Caffau:2011ik}. Although these hydrodynamical models of the solar atmosphere are much more realistic than the simple 1D atmospheres used by \citet{Grevesse:1998cy}, their results are not in agreement with the helioseismic determinations of the solar internal structure \citep[e.g.,][]{Basu:2008fo,ChristensenDalsgaard:2009et,Serenelli:2009ev}. Similarly, these simulations predict variations in the efficiency of convection as a function of $\teff$ and $\feh$ \citep[e.g.,][]{Trampedach:2014fo,2014arXiv1403.1062M}. In hydrostatic evolutionary models this process is mimicked by the mixing-length theory and its efficiency is represented by one parameter ($\alpha_\mathrm{MLT}$) calibrated to reproduce the properties of the Sun.

A thorough analysis of the virtues of 3D model predictions compared to those of 1D models goes beyond the scope of this paper, but we are interested in knowing the impact of these results on our fitted stellar properties. For the case of the solar abundances we have chosen to use the compilation of \citet{Asplund:2009eu} since its determinations represent an extreme case with respect to those of \citet{Grevesse:1998cy}, while the results of \citet{Caffau:2011ik} lie approximately in between these two determinations. In terms of convective efficiency our standard grid uses a value calibrated to the solar properties; to test the impact of only changing $\alpha_\mathrm{MLT}$ we have computed grids with the \citet{Grevesse:1998cy} abundances varying this parameter by $\pm$0.3, in agreement with the changes predicted by 3D simulations \citep[see e.g.,][]{Trampedach:2014fo}. The resulting sets of models (AS09, GS98al$+$, GS98al$-$) are summarised in Table~\ref{tab:grids}.

The stellar parameters obtained with BASTA and these three grids deviate systematically from the GS98sta results. When comparing to the values obtained with the \citet{Asplund:2009eu} solar abundances, the median and standard deviation of the fractional differences is $-0.1\pm0.3\%$ (density), $+0.3\pm0.3\%$ (radius), $+0.9\pm0.6\%$ (mass), and $+4.0\pm3.3\%$ (age). An increase in the mixing-length parameter on the other hand produces changes at the level of $+0.9\pm0.7\%$ (density), $-0.9\pm0.6\%$ (radius), $-2.4\pm2.2\%$ (mass), and$-0.2\pm9.0\%$ (age), and of opposite sign if the $\alpha_{\rm MLT}$ value is decreased. Given that these differences are always smaller than the statistical errors quoted in the previous section, we have not added them to the error budget given in Table~\ref{tab:stellar}. We consider the dispersions as the major contribution to the uncertainties, and add them in quadrature\footnote{Giving the scatter from changing the mixing-length half the weight as its contribution is symmetric} to quote a reference systematic effect arising from the difference in the input physics of 0.8\% (density), 0.7\% (radius), 2.3\% (mass), and 9.6\%(age).
\subsection{Systematic uncertainties from fitting methods}\label{ssec_meth}
As described in Sections~\ref{ssec_bayes} and~\ref{ssec_cen}, our recommended stellar properties are derived fitting the frequency ratios $\runo$ and $\rdos$ and the spectroscopic parameters to two chosen grids of models depending on the mass range. In many asteroseismic analyses it is customary to fit the individual modes frequencies (after applying a suitable surface correction) instead of frequency combinations. To isolate the impact of the choice of fitted seismic observables, we take advantage of the flexibility of BASTA in its input and compute stellar properties fitting the spectroscopic $\teff$ and $\feh$ and the individual oscillation frequencies (corrected with the \citet{Kjeldsen:2008kw} prescription). The median fractional differences in the results obtained with BASTA$+$GS98sta using frequencies and ratios is below 0.5\% in density and radius, 1\% in mass and 7\% in age. These values should be considered as a minimum set of systematic uncertainties based only on the choice of observables.

Applying the BASTA to one grid of GARSTEC models corresponds to a single determination of stellar properties at a fixed set of input physics, evolutionary code, pulsation code, and minimisation technique. Throughout the years several algorithms have been developed to reproduce asteroseismic data making use of a variety of stellar evolution and oscillation codes, considering different assumptions on the physics included and the treatment of the data to be fitted. We aim now at determining the systematic uncertainty arising from the use of different fitting pipelines, including the numerics of the evolution and pulsation codes as well as the minimisation routine and chosen data sets.

Bearing this in mind, we have obtained results for our sample of KOIs using three different methods: the ASTEC Fitting (ASTFIT) method, the YALE Monte-Carlo Method (YMCM), and the Asteroseismic Modeling Portal (AMP). To isolate the effects of using different codes and assumptions on data treatment each algorithm has been applied using input physics as similar as possible to the standard GARSTEC grid (GS98sta), i.e., not including the effects of microscopic diffusion nor overshooting. Thus, all comparisons in this Section are made between the results of these fitting algorithms and the BASTA$+$GS98sta combination. A description of the methods can be found in Appendix~\ref{app_meth}, but we repeat here the main differences in each of the implementations with respect to the BASTA$+$GS98sta.

The ASTFIT method uses the ASTEC evolutionary code \citep[][]{ChristensenDalsgaard:2008bi} and was applied with the same physics as in the GS98sta grid (including a solar-calibrated mixing-length and $\Delta Y/\Delta Z=1.4$). Its main difference with the BASTA are the asteroseismic variables to be fitted: instead of using frequency ratios, this algorithm matches the individual frequencies mimicking the surface effects via a scaled function determined from the frequency corrections found in the solar case (see Appendix~\ref{ssec_fit_jcd} for details).

The YMCM differs from the BASTA$+$GS98sta combination in its evolution code \citep[YREC,][]{Demarque:2007ij} and its pulsation code \citep[first used in][]{1994A&AS..107..421A}. Also, the assumptions on the initial composition are different: no helium to metal enrichment ratio is applied and the initial helium abundance is a free parameter in the minimisation process. The stellar properties are obtained from a weighted average of the fit to the individual frequencies (with a solar-function correction for surface effects, see Appendix~\ref{ssec_fit_bas}), the ratios, and the spectroscopic constraints. YMCM takes into account the correlations introduced by the construction of the frequency ratios (see equations~\ref{eqn:r02} and~\ref{eqn:rat}).

In the case of the AMP the evolution and pulsation code are the same as in ASTFIT \citep[][]{Metcalfe:2009ed}. However its standard configuration adjusts both the mixing-length parameter and initial abundances when searching for the best fit model. The latter is found using a weighted average between the fit to the individual frequencies (corrected with the \citet{Kjeldsen:2008kw} formulation), the frequency ratios $\runo$ and $\rdos$ and the spectroscopic constraints (see Appendix~\ref{ssec_fit_tra}). Correlations are not taken into account in the frequency combinations and these are assumed to be independent.
\begin{figure*}
\includegraphics[angle=90,scale=.70]{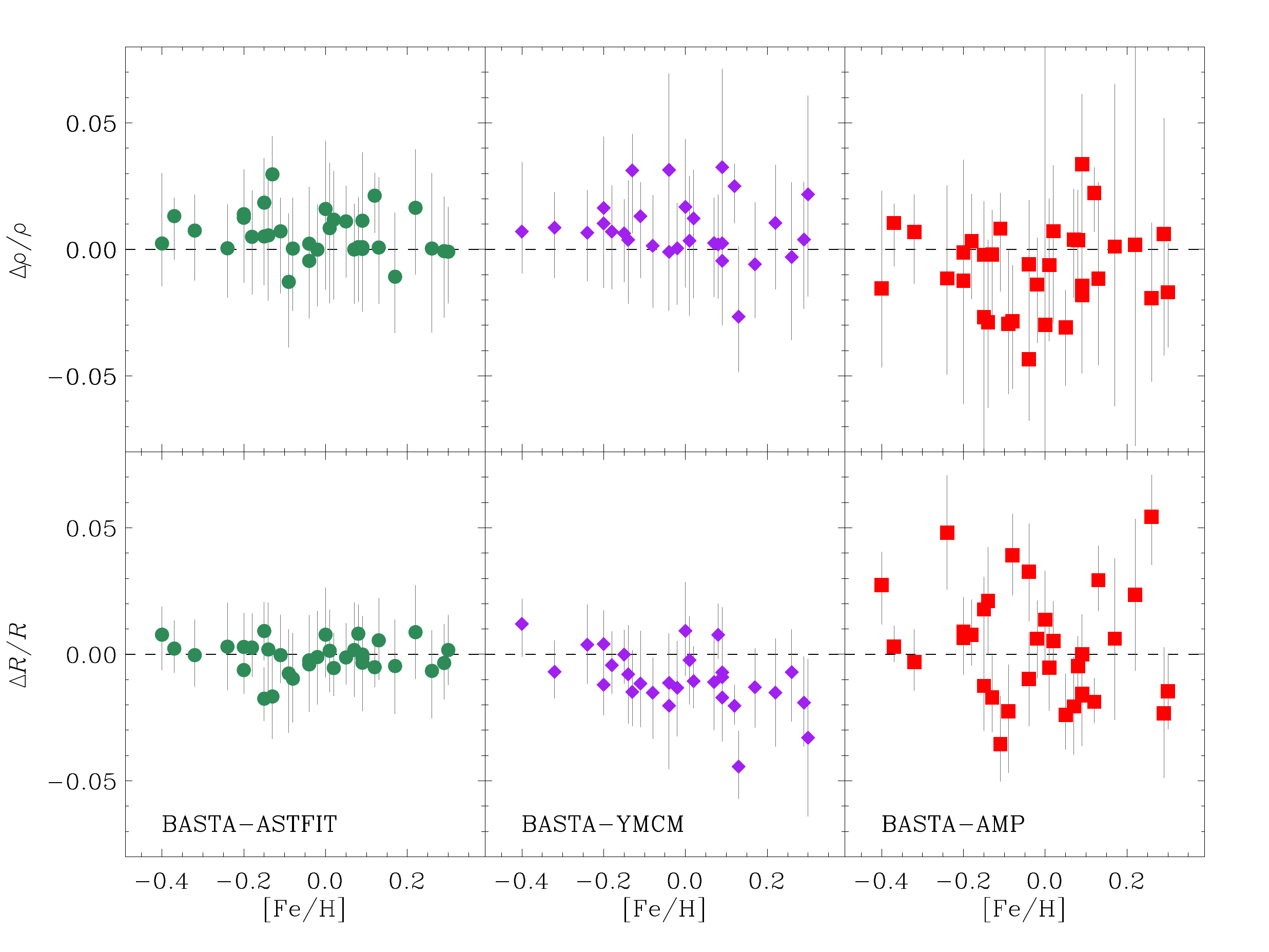}
\caption{Fractional differences in stellar density and radius for the three fitting methods compared to the BASTA results as a function of spectroscopic metallicity, given as (BASTA-other).}
\label{fig:fitting_rho}
\end{figure*}

Figure~\ref{fig:fitting_rho} shows the fractional differences obtained in mean density and radius from the different methods. It is worth noticing that the YMCM could not provide a  solution in four cases where the best fit was found for helium abundances below the standard Big Bang nucleosynthesis (SBBN) value of $Y_0=0.248$. There is overall good agreement in the results, with the AMP values showing a larger scatter than the other two methods. The closest matches are obtained between the BASTA and ASTFIT; not surprisingly since these are computed with the exact same input physics. Median differences and standard deviations in densities and radii are below the $\sim$1\% level, while the comparison with the YMCM and AMP leads to median differences of the same order but a larger scatter ($\sim$2\%). This is a remarkable level of agreement considering the variety of evolutionary and pulsation codes used as well as the treatment of asteroseismic data. 
\begin{figure*}
\includegraphics[angle=90,scale=.70]{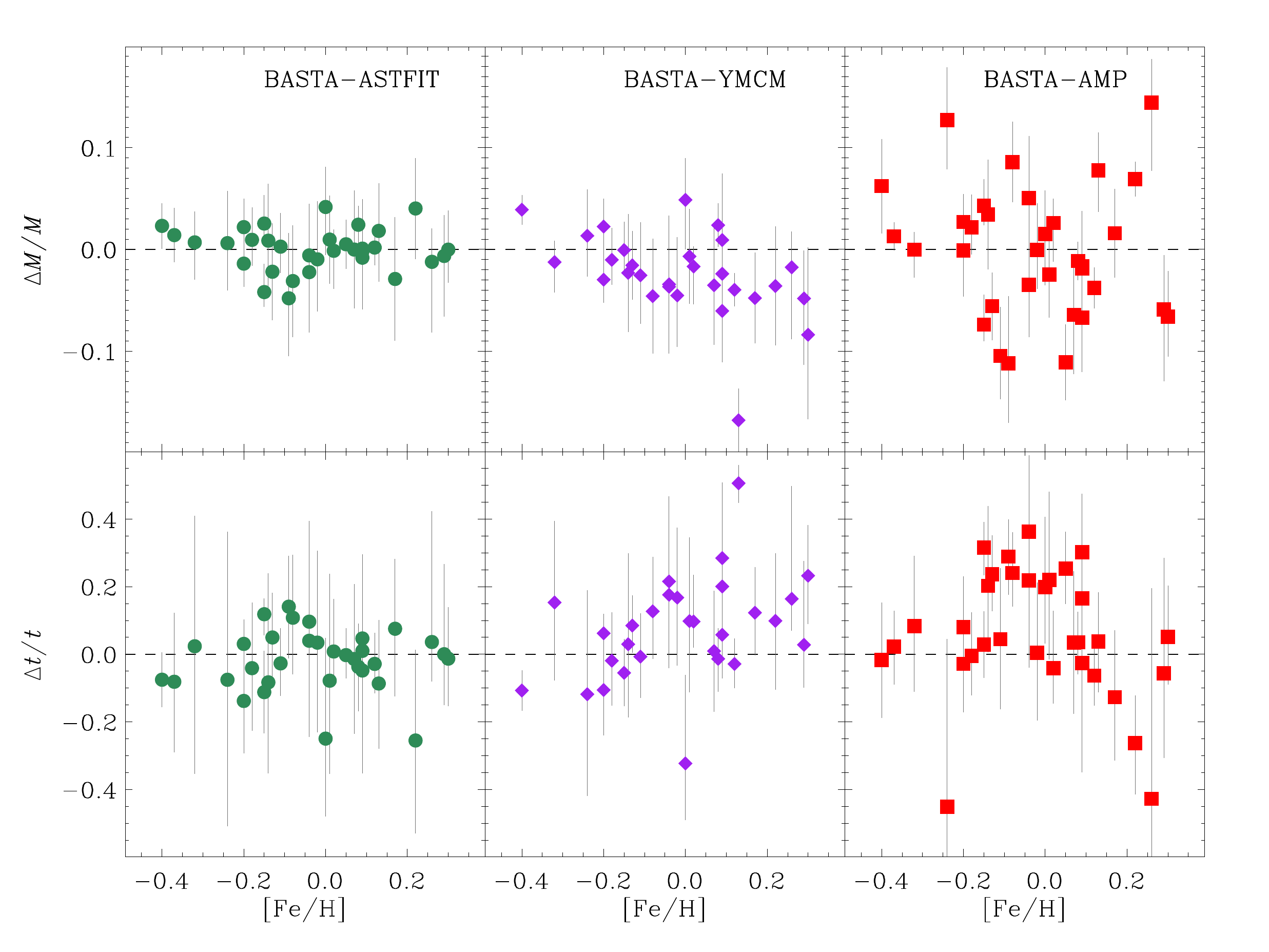}
\caption{Fractional differences in stellar mass and age for the three fitting methods compared to the BASTA results as a function of spectroscopic metallicity, given as (BASTA-other).}\label{fig:fitting_mass}
\end{figure*}

There is evidence of a metallicity dependence in the density and radius differences between the BASTA and YMCM results, while the comparison with the AMP does not suggest any correlation with the spectroscopic parameters. Similar features can be seen in Fig.~\ref{fig:fitting_mass}, where the fractional differences in mass and age are shown. Once again there is excellent agreement between the BASTA and ASTFIT, with median and standard deviations in mass and age of $0.8\pm2$\% and $1.2\pm9$\% respectively and within the uncertainties in practically all cases. The previous trends with spectroscopic parameters remain with respect to the YMCM and AMP, namely a metallicity dependence in the first case and seemingly random scatter in the second one. This leads to mass estimates differing by up to $\sim$15\% and, via the mass-age correlation, to corresponding age differences of more than 40\%.

While the origin of these discrepancies will be explored in the following section, for the sake of the current analysis we use the scatter in the comparison between the BASTA and the ASTFIT results as a measurement of the systematic uncertainty arising purely from evolutionary codes and fitting methods, which conservatively accounts for 1\% in density and radius, 2\% in mass, and 9\% in age. This level of agreement is comparable to the one obtained with BASTA$+$GS98sta when changing the observables from ratios to individual frequencies, showing that remarkably precise determinations of stellar properties are possible from asteroseismology despite the use of different evolutionary codes as long as the considered input physics is kept as similar as possible.
\subsection{Systematic uncertainties from the initial helium abundance}\label{ssec_helium}
\begin{figure*}
\includegraphics[angle=90,scale=.70]{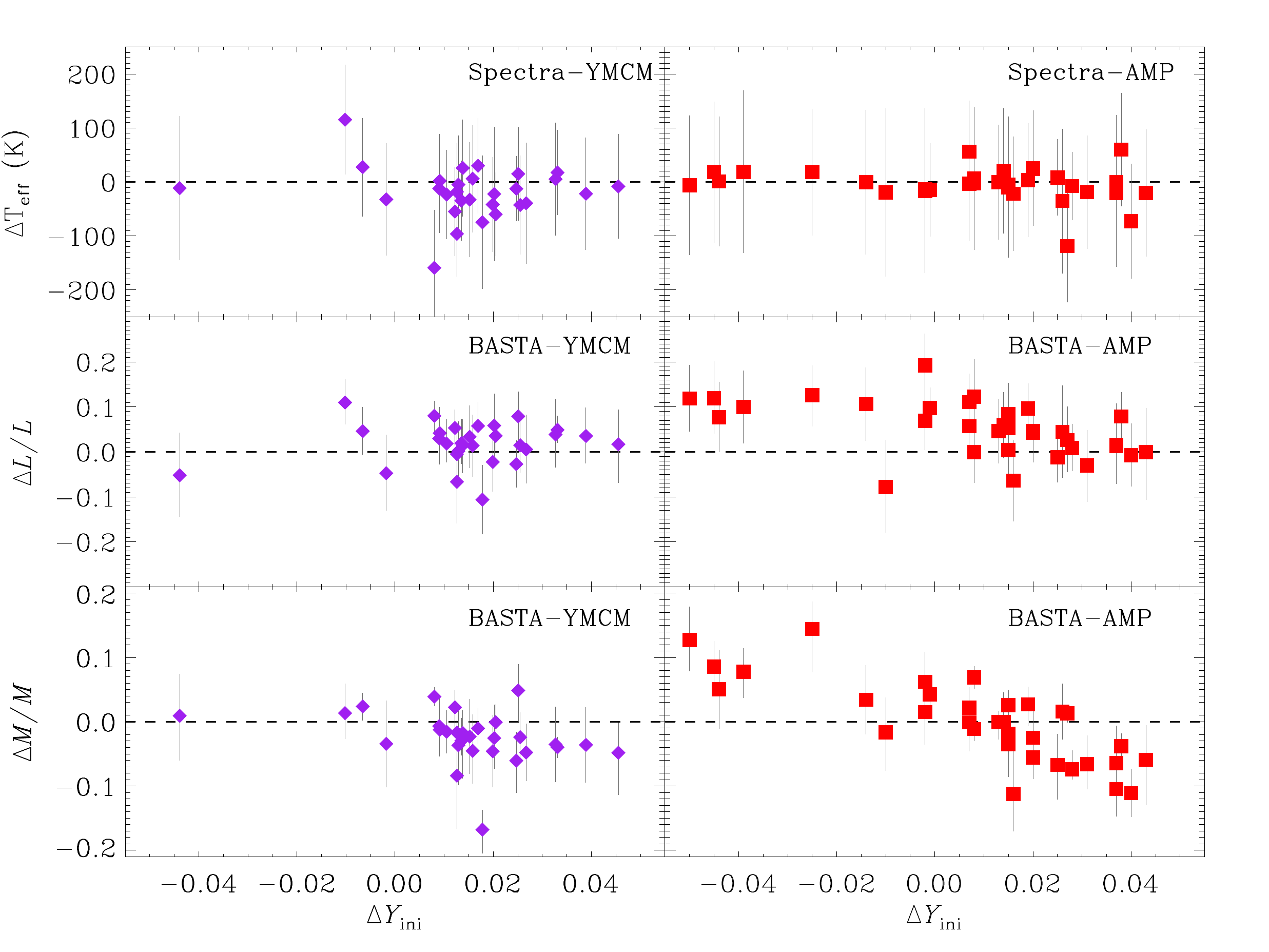}
\caption{Comparison of stellar properties as a function of the difference in the initial helium abundance, given as (BASTA-code). {\it Top panels}: difference in effective temperature between the spectroscopically measured one and that extracted from the fitting methods. {\it Middle panels}: fractional differences in the luminosities predicted by the codes. {\it Bottom panels}: Fractional mass differences between codes. See text for details.}
\label{fig:Yini_codes}
\end{figure*}

To understand now the origin of the discrepancies between YMCM, AMP, and the BASTA$+$GS98sta combination highlighted in the previous section, we naturally focus on the parameters that have been treated differently in the algorithms. The bottom panels in Fig.~\ref{fig:Yini_codes} show the fractional mass difference as a function of the difference in the initial helium abundance. In both cases a correlation between $\Delta M/M$ and $\Delta Y_\mathrm{ini}$ can be seen, with higher masses predicted by the YMCM and AMP for lower values of the initial helium fraction. A similar trend can be seen when comparing the luminosities (middle panels in Fig.~\ref{fig:Yini_codes}), which turns out to be a consequence of the different weights each fitting method gives to the spectroscopic constraints (see Appendix~\ref{app_meth} and section~\ref{ssec_bayes} for details).

Since there is good agreement in the median density and radii determined by all pipelines (better than 2\% with a larger scatter for AMP), differences in luminosity arise from differences in effective temperature of each model. The top panels of Fig.~\ref{fig:Yini_codes} show the match to the spectroscopic $\teff$ value predicted by each algorithm, revealing a very close agreement between these codes and the spectra (mean difference of 20~K and 35~K for AMP and YMCM, respectively). ASTFIT and the BASTA on the other hand predict $\teff$ values systematically higher than the spectroscopic ones at a level of $\sim$45~K and 70~K, respectively. The reason for this difference is that the AMP and YMCM have the flexibility to fit $\teff$ at a fixed radius by changing the luminosity via its strong dependence on the molecular weight and thus the helium abundance. Our final results from the BASTA given in Table~\ref{tab:stellar} do not suffer from this systematic offset in $\teff$ as they include microscopic diffusion which effectively compensates for this effect as shown in Fig.~\ref{fig:koi69_diff} (see also discussion in section~\ref{ssec_cen}).

The initial helium abundance in low-and intermediate-mass stars is a poorly constrained quantity as helium lines are not detectable in their spectra. Usually these abundances are computed assuming a linear enrichment law $\Delta Y/\Delta Z$ anchored to primordial values $(Y_0,Z_0)$ determined from SBBN or a solar calibration. The actual slope of the relation has been determined from different sources such as low-main-sequence stars in the solar neighbourhood \citep{Jimenez:2003il, Casagrande:2007ck}, galactic and extragalactic H-II regions \citep{2006AJ....132.2326B}, and stellar evolution theory for a given initial mass function \citep[e.g.,][]{Chiappini:2002dk}. There is no consensus yet on the relation being linear across all metallicity values or the slope changing for different types of stars and metallicity ranges. However, there is broad agreement on the expected range for the relation being between $1\le\Delta Y/\Delta Z\le3$ \citep[except for some very peculiar objects such as $\omega$~Centauri, see][]{2004ApJ...612L..25N}. Depending on the chosen solar composition, determination of the initial helium abundance of the Sun agrees with a slope of $1.7\le\Delta Y/\Delta Z\le2.2$ \citep[see][]{Serenelli:2010gu}.
\begin{figure*}
\includegraphics[width=84mm]{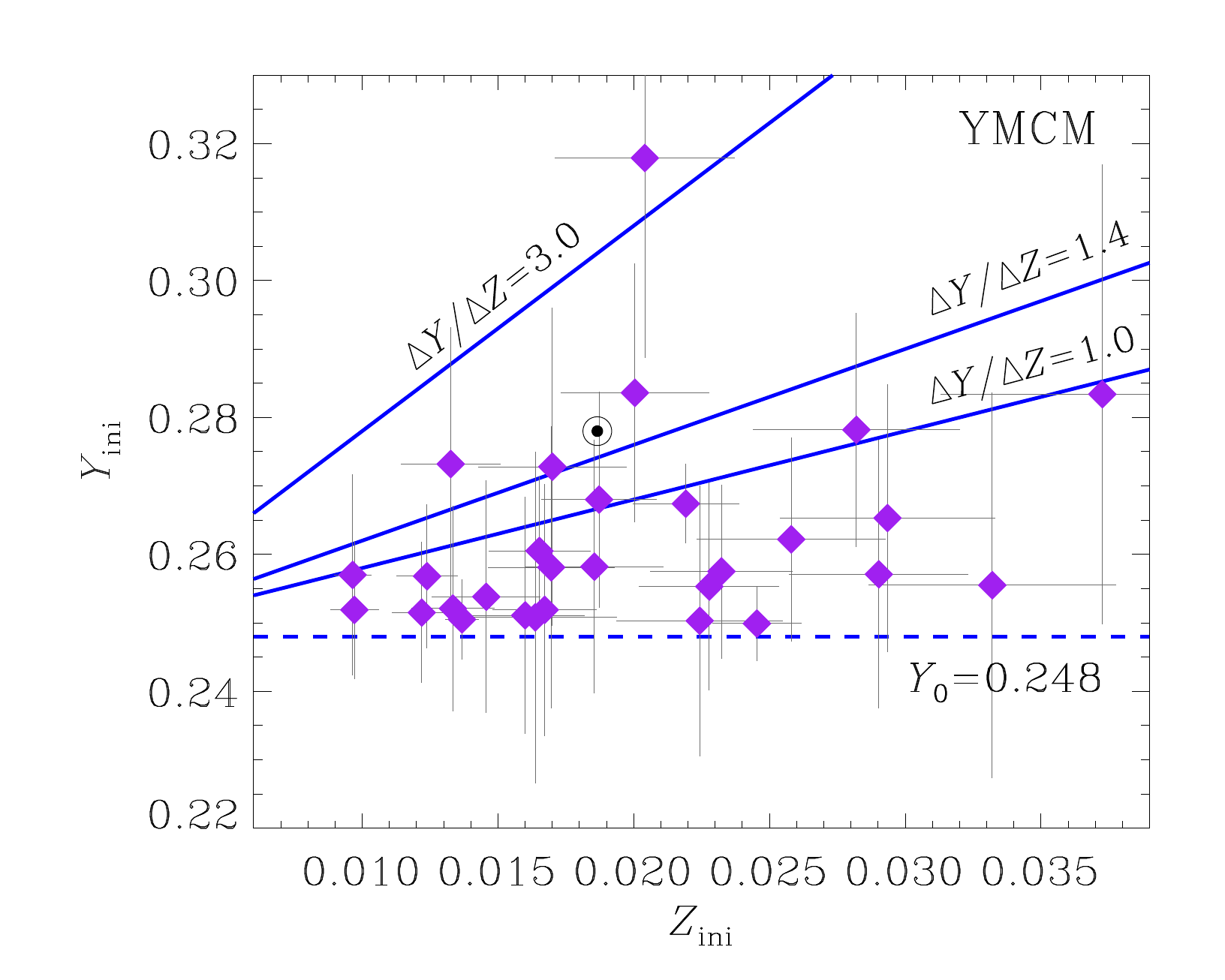}
\includegraphics[width=84mm]{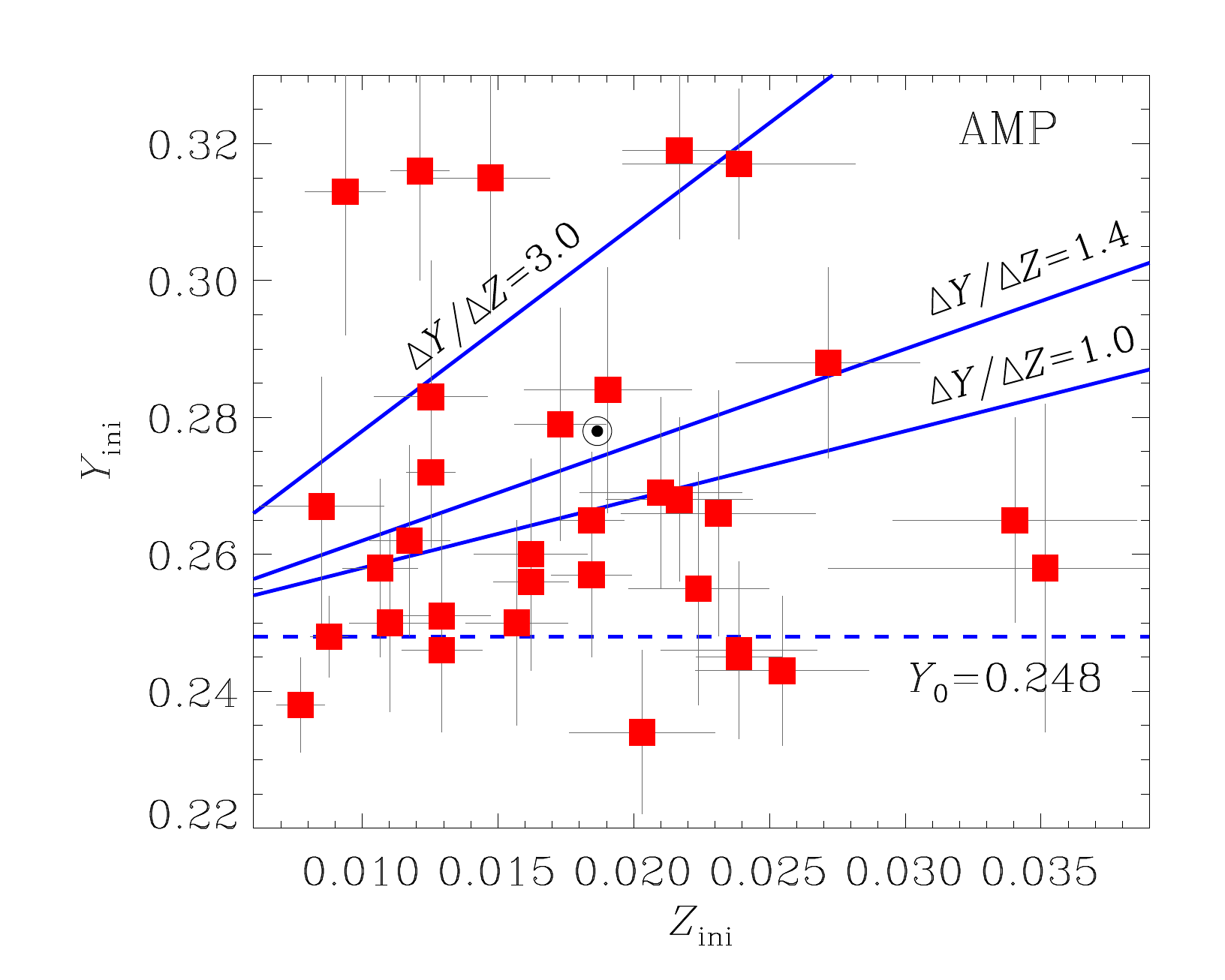}
\caption{Initial metallicity versus initial helium abundance for the results determined by two different codes. Horizontal dashed lines shows the primordial helium abundance predicted by SBBN, while solid lines depict different galactic enrichment laws anchored to $Y_0=0.248$ and $Z_0=0$. Solar symbol depicts the initial solar abundance from \citet{Serenelli:2010gu}. {\it Left}: YMCM. {\it Right}: AMP. See text for details.}\label{fig:dydz}
\end{figure*}

Figure~\ref{fig:dydz} compares the predictions of helium to metal enrichment laws of different slopes with the initial abundances determined by the YMCM and AMP. We remind the reader that the BASTA and ASTFIT results are anchored to $\Delta Y/\Delta Z=1.4$ and therefore follow a straight line if plotted in the figure. As expected, a large majority of the YMCM results lie below this line and in many cases very close to the primordial SBBN value which was used as a hard edge in the modelling procedure. The AMP results on the other hand only force a hard cut in $Y_\mathrm{ini}=0.22$ and show a much larger scatter as well as five targets with initial helium abundances below the SBBN line. This occurs for stars both at low metallicities and at and above the solar value ($Z_\mathrm{ini}\sim0.018$). The AMP also predicts high initial helium for stars of metallicity lower than the Sun, which occurs only for one case in the YMCM results.

Only two of our targets have available {\it Hipparcos} parallaxes from \citet{VanLeeuwen:2007dc}: Kepler-21 ($\pi=8.86\pm0.58$~mas) and KOI-3158 ($\pi=28.03\pm0.82$~mas). Using the asteroseismic $\logg$ value and spectroscopic metallicity, we applied the InfraRed Flux Method (see section~\ref{ssec_dis}) to determine the bolometric flux and the luminosity of the stars. The resulting values of $L/L_\odot=5.54\pm0.78$ (Kepler-21) and $L/L_\odot=0.38\pm0.03$ (KOI-3158) are compatible with those determined with the BASTA (see Table~\ref{tab:stellar}). ASTFIT predicts the correct luminosity for KOI-3128 ($L/L_\odot=0.35\pm0.02$) while its determination for Kepler-21 results in a slightly lower value than observed ($L/L_\odot=4.59\pm0.11$). The reason for this discrepancy is that ASTFIT favours the low-mass solution in the bimodal mass distribution observed for Kepler-21 (see section~\ref{ssec_cen}); the final results of the BASTA are based on the overshooting grid that favours a higher-mass value. For comparison the low-mass solution obtained with the BASTA+GS98sta predicts a luminosity of $L/L_\odot=4.80^{+0.12}_{-0.10}$.

The results from AMP return similar luminosities as ASTFIT: $L/L_\odot=0.36\pm0.02$ (KOI-3158) and $L/L_\odot=4.37\pm0.19$ (Kepler-21). However, in both cases the $Y_{\rm ini}$ values from AMP are slightly below the SBBN one: the luminosity of KOI-3158 is still well reproduced because at such low metallicities the initial helium abundance at $\Delta Y/\Delta Z=1.4$ is not very different from the primordial one ($Y_{\rm ini}\sim0.258$). The resulting mass of Kepler-21 is higher than that from ASTFIT but the luminosity is lower as a consequence of the difference in initial helium abundance at compositions closer to solar ($Y_{\rm ini}\sim0.270$).

The correlation between mass and helium abundance severely affects asteroseismically derived stellar properties \citep[see section 4.3 in][]{Lebreton:2014gf}, and has recurrently appeared in studies fitting individual frequencies \citep[e.g.,][]{Mathur:2012bj,Metcalfe:2014ig}. Assessing if the sub-SBBN results have astrophysical meaning goes beyond the scope of this paper, but the dependence on helium abundance of asteroseismic solutions needs to be further investigated and thoroughly understood. In order to estimate an uncertainty arising from the changes in initial helium content, we compare the results of BASTA, YMCM, and AMP for the subset of targets falling within the expected range of $1\le\Delta Y/\Delta Z\le3$. There are five and nine targets fulfilling this criterion in the YMCM and AMP results, respectively (see Fig.~\ref{fig:dydz}). The average standard deviations around the median of the results are 1.6\% (radius), 1.7\% (density), 3.6\% (mass), and 16.8\% (age). These uncertainties are comparable to the statistical ones obtained with BASTA$+$GARSTEC grids, as described at the end of section~\ref{ssec_cen}.

In an initial attempt to solving the issue of asteroseismic analysis favouring sub-SBBN helium models, we consider the correlation of properties with initial helium abundances shown in Fig.~\ref{fig:Yini_codes}. Our results for the two targets where parallax-derived luminosities are available suggest a better agreement with grids computed assuming $\Delta Y/\Delta Z=1.4$, although the uncertainties are still rather large and AMP results are only slightly more than $1\,\sigma$ away in Kepler-21. We expect much tighter observational constraints on the luminosity from the \emph{Gaia} mission that will essentially translate into a constraint on $Y_\mathrm{ini}$, breaking the degeneracy and subjecting the $\Delta Y/\Delta Z$ relation to an additional test from asteroseismology.
\subsection{Distances}\label{ssec_dis}
We have determined distances to our targets combining the results of the InfraRed Flux Method \citep[][]{Casagrande:2010hj} with asteroseismically determined stellar parameters \citep[][]{SilvaAguirre:2011es}. This technique has been shown to provide accurate {\it Hipparcos}-quality distances for individual targets \citep[][]{SilvaAguirre:2012du} and has been recently applied to the first Galactic Archaeology study of the {\it Kepler} field \citep[][]{Casagrande:2014bd}.

In its current implementation the method relies on multi-band {\it griz} photometry from the original Kepler Input Catalogue \citep[KIC,][]{Brown:2011dr}, $JHK_{\rm S}$ magnitudes from 2MASS, and the spectroscopically determined metallicity to recover the bolometric flux as a function of surface gravity. The precise $\logg$ values obtained from asteroseismology are included in an iterative process until convergence in reddening and effective temperature is achieved \citep[see section~3 in][for details]{SilvaAguirre:2012du}.

Unfortunately only 22 of our 33 targets have reliable KIC photometry taken in all {\it griz} filters, and we have initially determined distances for this subsample using the combination of the IRFM and asteroseismology. With the aim of providing a complete and homogeneous set of stellar distances for the full cohort, we have used the $J-K_{\rm S}$ angular diameter calibration of \citet{Casagrande:2006cb} to obtain distances to all targets, and compared them to the full IRFM results. For the 22 targets, the median and standard deviation of the fractional distance difference between both determinations is $0.3\pm5.1$\%, showing that the calibration works well for the selected sample. The distances given in Table~\ref{tab:stellar} for all targets have been determined from the colour-angular diameter calibration, and have a fractional median uncertainty of 4.4\%.

The two stars in our sample with measured {\it Hipparcos} parallaxes allow for an independent check on our distances. The resulting values of $d=112.9\pm7.4$~pc for Kepler-21 and $d=35.7\pm1.04$~pc for KOI-3158 are compatible within their uncertainties to those determined by our method (114~pc and 33~pc respectively, see Table~\ref{tab:stellar}).
\section{Comparison with asteroseismic scaling relations}\label{sec_scal}
The analysis presented in the previous sections has been possible for a subsample of almost half the currently studied KOIs where asteroseismic data are available. In the rest, SNRs are too low or the frequency resolution not sufficiently high to robustly extract individual oscillation frequencies \citep[see e.g.,][]{Chaplin:2014jf}. Under these circumstances, asteroseismic determination of stellar parameters is carried out based on the properties of the readily extractable average global parameters $\langle\dnu\rangle$ and $\num$. The former quantity is a proxy of the mean stellar density $\bar{\rho}$ \citep{Ulrich:1986ge}, while the latter has been shown to scale with the surface gravity and effective temperature \citep{Brown:1991cv,Kjeldsen:1995tr,Bedding:2003dz}:
\begin{equation}\label{eqn:sca_dnu} 
\left(\frac{\langle\dnu\rangle}{\langle\dnu_\odot\rangle}\right)^{2} \simeq \frac{\bar{\rho}}{\bar{\rho}_\odot}\,, 
\end{equation}
\begin{equation}\label{eqn:sca_num} 
\frac{\num}{\nu_{\mathrm{max},\odot}} \simeq \frac{M}{\msun}{\left(\frac{R}{{\rm R_\odot}}\right)^{-2}} \left(\frac{\teff}{T_{\mathrm{eff},\odot}}\right)^{-1/2}\,. 
\end{equation}

Here, $\bar{\rho}_\odot$ and $T_{\mathrm{eff},\odot}$ correspond to the solar properties. Consequently, $\langle\dnu\rangle$ and $\num$ are the basis of the asteroseismic scaling relations
\begin{equation}\label{eqn:sca_mass} 
\frac{M}{\msun} \simeq \left(\frac{\num}{\nu_{\mathrm{max},\odot}}\right)^{3} \left(\frac{\langle\dnu\rangle}{\langle\dnu_\odot\rangle}\right)^{-4}\left(\frac{\teff}{T_{\mathrm{eff},\odot}}\right)^{3/2}\,, 
\end{equation}
\begin{equation}\label{eqn:sca_rad} 
\frac{R}{{\rm R_\odot}} \simeq \left(\frac{\num}{\nu_{\mathrm{max},\odot}}\right) \left(\frac{\langle\dnu\rangle}{\langle\dnu_\odot\rangle}\right)^{-2}\left(\frac{\teff}{T_{\mathrm{eff},\odot}}\right)^{1/2}\,, 
\end{equation}
where $\langle\dnu_\odot\rangle$ and $\nu_{\mathrm{max},\odot}$ are the solar values as determined by the same method used to analyse the data.

When a determination of $\teff$ is available, equations~\ref{eqn:sca_mass}~and~\ref{eqn:sca_rad} provide masses and radii independent of any stellar models \citep[e.g.,][]{Stello:2008gt,SilvaAguirre:2011es}. Coupled to evolutionary tracks the asteroseismic scaling relations can also provide an age estimation via the so-called grid-based method, that matches the spectroscopic constraints and measured $\langle\dnu\rangle$ and $\num$ to the same values predicted from stellar models \citep[e.g.,][]{Basu:2010hv,Gai:2011it,SilvaAguirre:2014fj,MLundkvist:2014gm}. These techniques have been extensively used in characterisation of the CoRoT and {\it Kepler} samples \citep[e.g.,][]{Mosser:2010gx,Huber:2014dh}, studies of open clusters \citep[e.g.,][]{Basu:2011cc,Stello:2011hu}, galactic archaeology \citep[e.g.,][]{Miglio:2013hh,Casagrande:2014bd}, and will form the basis of asteroseismic analysis for the vast majority of targets observed by the repurposed Kepler mission \citep[K2,][]{Howell:2014ju}.

The $\langle\dnu\rangle$ relation in equation~\ref{eqn:sca_dnu} has direct theoretical foundation while the $\num$ scaling in equation~\ref{eqn:sca_num} is mostly an empirical relation \citep[although see][]{Belkacem:2011hm}. For these reasons the asteroseismic scaling relations need to be thoroughly tested and calibrated, and have been the subject of scrutiny in the past few years. Studies have shown that radii of main-sequence stars predicted from asteroseismology agree with determinations from parallaxes and interferometry \citep[e.g.,][]{North:2007hl,Bedding:2010bk,2012rgps.book...11M,SilvaAguirre:2012du,Huber:2012iv,White:2013bu}, while asteroseismic masses still remain to be confirmed by independent measurements. In the case of red giant stars recent work has cautioned about the predictions from scaling relations in first ascent red giant and red clump stars \citep[][]{Miglio:2012dm}, expected masses of metal-poor stars \citep[][]{Epstein:2014ct}, and open cluster properties of main-sequence binaries \citep[][]{Brogaard:2012bo,Miglio:2012dm}. Results from detailed modelling also suggest that the scaling relations underestimate the mean stellar density of red giant stars by $\sim$5\% \citep[][]{Huber:2013ha}. 

While acknowledging the shortcomings of our current understanding of the physics behind equations~\ref{eqn:sca_dnu}~and~\ref{eqn:sca_num} we note that the discrepancies arise in regimes of metallicity and evolutionary state far from the main sequence for middle-aged stars near the mass of our Sun from where these relations are extrapolated. Our sample of KOIs presents an opportunity to compare determinations from individual frequencies with those of the grid-based method in a parameter space closer to the Sun, and thus to validate the results of the scaling relations for population studies in this region of the Hertzsprung-Russell diagram. The BASTA presented in Section~\ref{ssec_bayes} is flexible and can be adapted to any set of input data. We include as observational constraints the quantities $\teff$, $\feh$, $\langle\dnu\rangle$ and $\num$ of each target, which have been determined by \citet{Huber:2013jb}. To derive their stellar properties, we need to determine $\num$ and $\langle\dnu\rangle$ in stellar evolution models.

The theoretical $\num$ value can only be obtained using equation~\ref{eqn:sca_num}, and we adopt for the reference solar value entering this equation $\nu_{\mathrm{max},\odot}=3090\pm30~\mu$Hz given by \citet{Huber:2011be}. The theoretical $\langle\dnu\rangle$ on the other hand can be obtained either from equation~\ref{eqn:sca_dnu} or directly from the individual frequencies of oscillations calculated for each stellar model. To implement the latter approach we must take into account the way $\langle\dnu\rangle$ is estimated from the observed frequencies and treat models and data as similarly as possible \citep[see][]{White:2011fw}. With this in mind, we adopt the following procedure: we determine the theoretical $\langle\dnu\rangle$ using a Gaussian-weighted linear fit to the $\ell=0$ modes of each model as a function of radial order $n$, centred at the $\num$ value obtained from equation~\ref{eqn:sca_num} and with a width of 0.25~$\num$.
\begin{figure}
\begin{center}
\includegraphics[width=84mm]{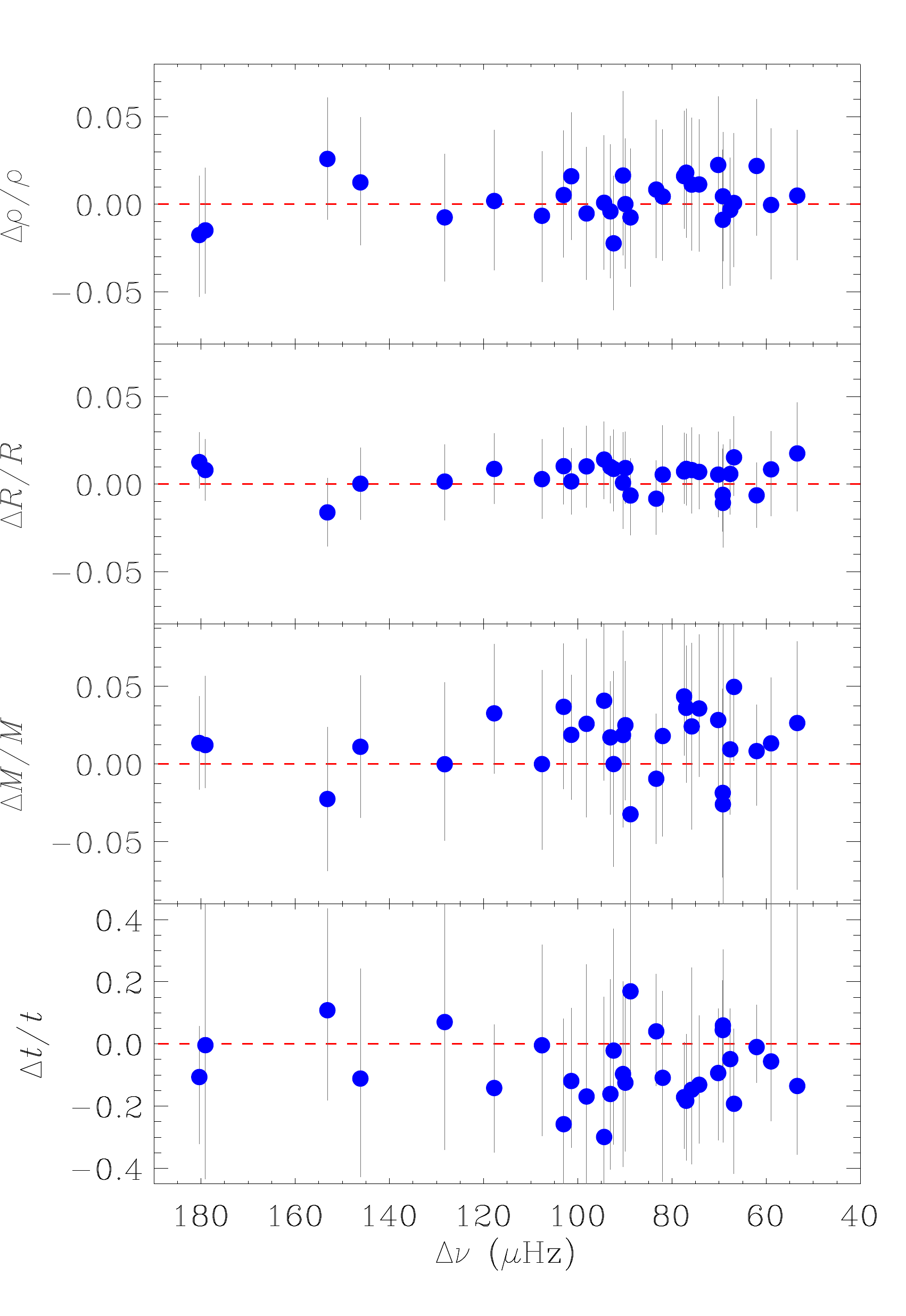}
\caption{Fractional differences in stellar properties determined from the frequency combinations and the scaling relations as a function of the large frequency separation, in the sense (combinations-scaling).}
\label{fig:scal_rels}
\end{center}
\end{figure}

Figure~\ref{fig:scal_rels} shows the comparison between the stellar properties determined with the BASTA in section~\ref{sec_res} using the frequency ratios and those obtained using the global asteroseismic parameters, where $\langle\dnu\rangle$ in the models has been computed using the aforementioned Gaussian fit to the theoretical frequencies. There are three outliers (same as in section~\ref{ssec_ove}, not shown in this figure) corresponding to stars presenting bimodal PDFs. For the rest of the sample there is excellent agreement between both determinations, with median fractional differences and standard deviations of $0.5\pm1.2$\% (density), $0.7\pm0.8$\% (radius), $1.8\pm2.1$\% (mass), and $-10.6\pm10.8$\% (age). These results are reassuring considering that median uncertainties in grid-based determinations are usually of the order of 2.2\% in radius, 2.8\% in density, 5.4\% in mass, and 25\% in age \citep{Chaplin:2014jf}. Thus, stellar properties in the $\teff$ and $\feh$ regime explored obtained from the global average asteroseismic parameters are compatible with those from individual frequency determinations and can be used when the latter are not available.

Finally, we have computed a set of stellar properties for our KOI sample using the two scaling relations (equations~\ref{eqn:sca_dnu} and ~\ref{eqn:sca_num}) to determine the global asteroseismic parameters, and the \citet{Huber:2011be} reference solar values ($\langle\dnu_\odot\rangle=135.1\pm0.1~\mu$Hz, $\nu_{\mathrm{max},\odot}=3090\pm30~\mu$Hz). The median fractional differences and standard deviations with the results obtained from frequency combinations are $-1.7\pm1.2$\% (density), $1.5\pm0.9$\% (radius), $2.6\pm2.2$\% (mass), and $-10.4\pm11.5$\% (age). The level of agreement is slightly worse than that obtained when the theoretical $\langle\dnu\rangle$ was computed from individual frequencies, but still well within the median uncertainties of the method quoted by \citet{Chaplin:2014jf}. However, compared to the results from the frequency ratios, the differences are of systematic nature with the scaling relations overestimating the density and underestimating the radius and mass. Departures from the $\langle\dnu\rangle$ scaling relation (equation~\ref{eqn:sca_dnu}) in stellar models can introduce biases in the results, and have already been reported by \citet{White:2011fw} and \citet{Chaplin:2014jf} in the main-sequence phase. A detailed exploration of this systematic biases goes beyond the scope of this study and will be addressed in a subsequent publication.
\section{Age correlations with exoplanet properties}\label{sec_exo}
\begin{figure}
\begin{center}
\includegraphics[width=84mm]{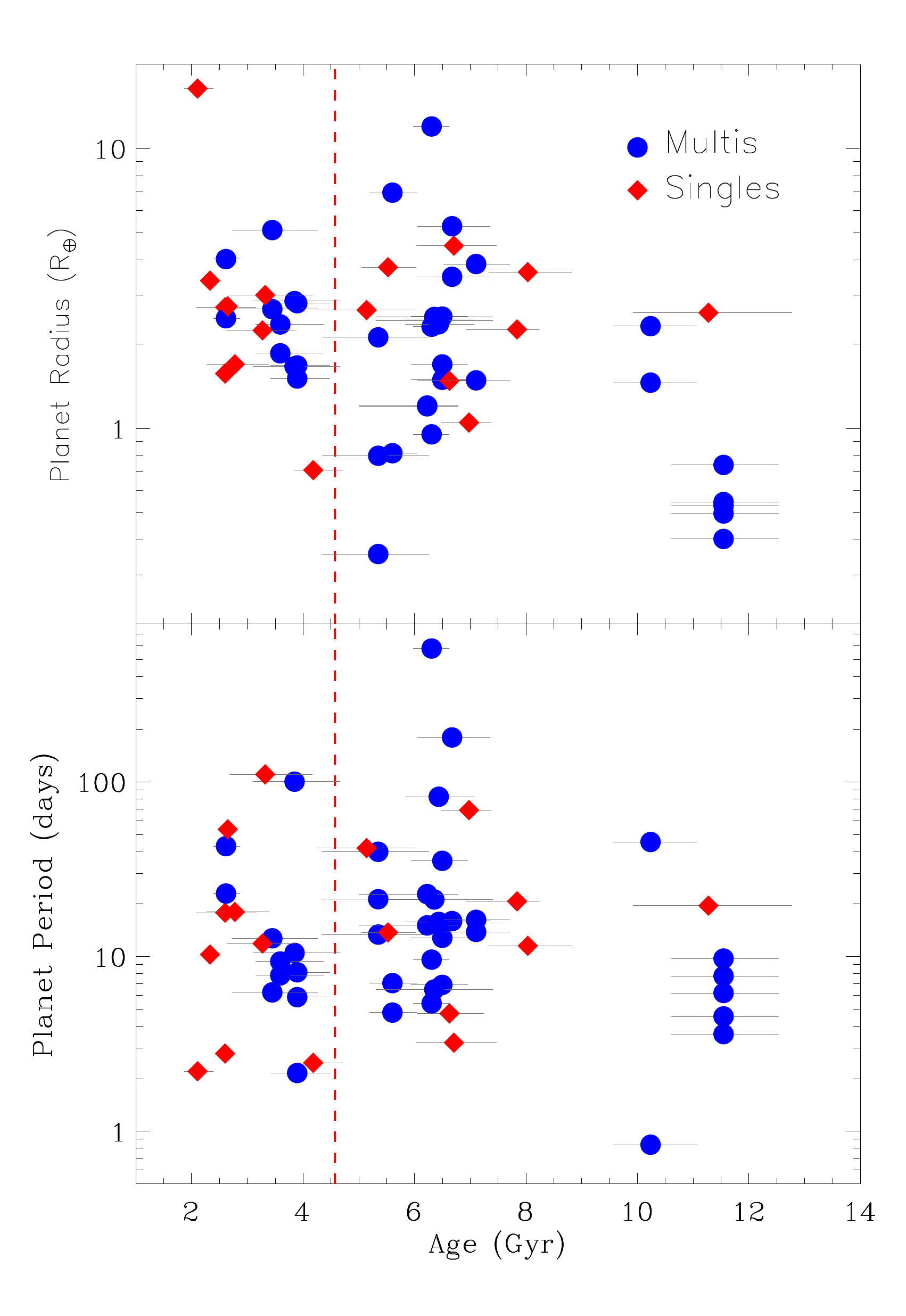}
\caption{Properties of single (red diamonds) and multiple (blue circles) systems as a function of stellar age. {\it Top panel:} planetary radius. {\it Bottom panel:} planetary period. The oldest multiple system correspond to the five planets detected around KOI-3158. Vertical dashed line marks the solar.}
\label{fig:exo_age}
\end{center}
\end{figure}

Using the stellar properties determined with the BASTA and shown in Table~\ref{tab:stellar}, we can investigate the distribution of planetary properties such as orbital period, radius and multiplicity as a function of stellar age. Figure~\ref{fig:exo_age} illustrates the age distribution of our sample as a function of planet radius and orbital period, showing that the majority of the high SNR asteroseismic host-stars are older than the Sun. Also, two main populations can be distinguished in age, at $\sim$3~Gyr and $\sim$6~Gyr. These features arise due to biases in the detection of stellar pulsations, whose amplitudes are proportional to the stellar luminosity \citep[e.g.,][]{Kjeldsen:1995tr}. The net result is that detections are favoured in F-type (with ages about 2-3 Gyr) and old G-type (i.e., 6 Gyr) stars that have high enough luminosities \citep[see e.g., Fig.~1 in][]{SilvaAguirre:2011es}. Given these age biases for high signal-to-noise asteroseismic detections, it should not be inferred that our finding of an average age greater than solar applies to either the {\it Kepler} exoplanet host star sample in general, nor the larger set of all Kepler dwarf stars comprising the exoplanet search program. However, it is interesting to look for age correlations using this sample and explore the future prospects for larger samples of asteroseismically determined stellar properties of host stars.

Figure~\ref{fig:exo_age} shows no obvious trends of stellar age with planet radii or orbital periods. This is not surprising given that evolutionary effects which are expected to affect planet radii or orbital periods such as photoevaporation and orbital migration are expected to occur within $<$1 Gyr \citep[][]{1996Natur.380..606L,Lopez:2013gh}, and our sample does not contain such young stars. Nevertheless, our sample suggests that the period and radius distribution appears to be approximately constant for ages $>$2 Gyr, and hence is likely representative for Sun-like host stars observed by {\it Kepler}. We also do not detect any significant difference in the multiplicity fraction between young ($\sim$2-4 Gyr) and old ($>$5 Gyr) stars.

\begin{figure}
\begin{center}
\includegraphics[width=84mm]{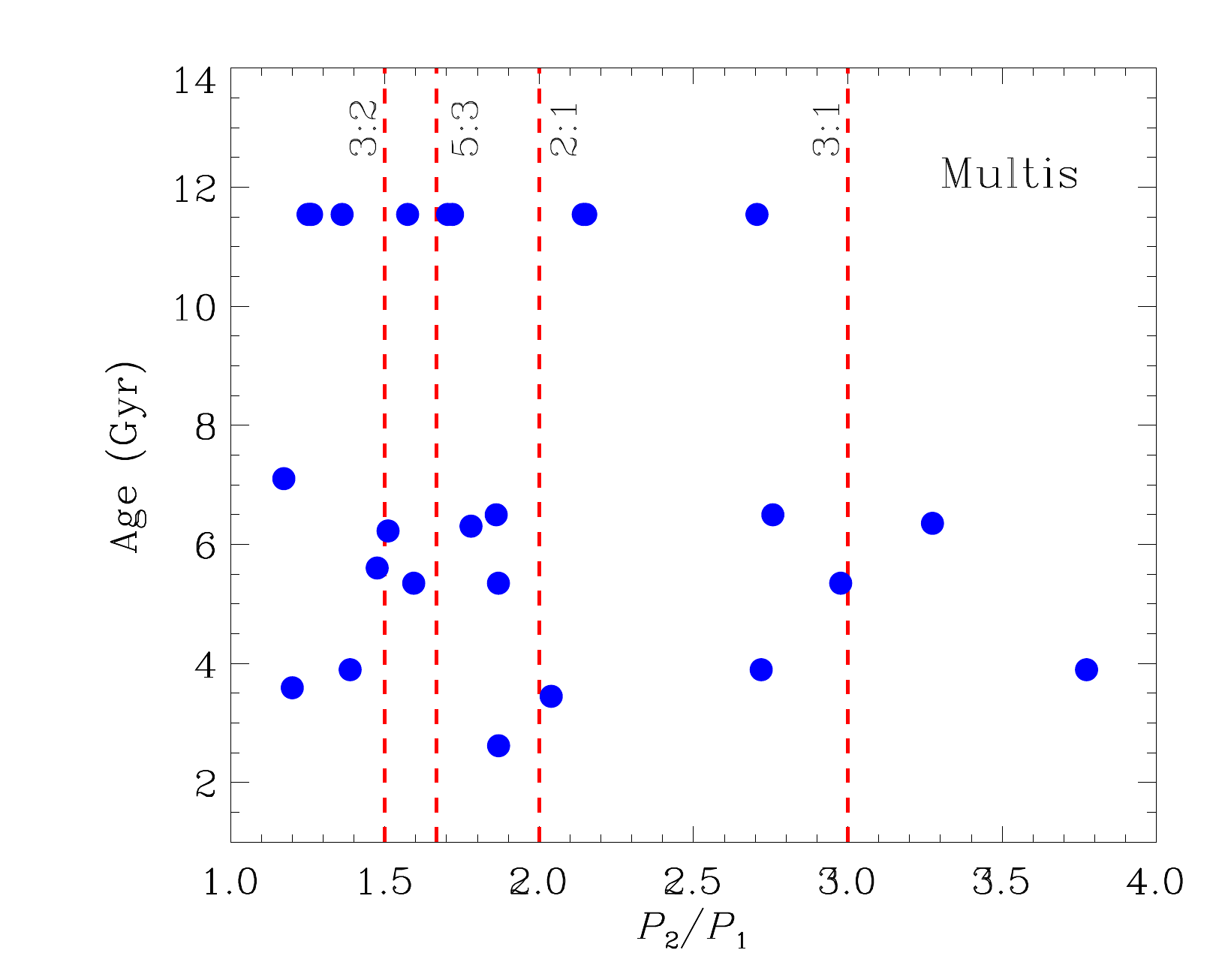}
\caption{Period ratios for the multiple systems in our sample. Dashed lines mark the 3$:$2, 5$:$3, 2$:$1, and 3$:$1 mean-motion resonances.}
\label{fig:exo_reson}
\end{center}
\end{figure}

Finally, Fig.~\ref{fig:exo_reson} shows the ratio of planetary periods for all known multiple systems in our sample. Studies of the larger Kepler planet-candidate sample have shown that most planets are not on exact resonant orbits, but that their period ratios tend to be near first-order resonances \citep[e.g.,][]{Fabrycky:2014hl}. Although our sample is small, the results seem to indicate that these conclusions hold independently of the age of the host stars.
\section{Conclusions}\label{sec_conc}
We have made use of the full time-base of observations from the {\it Kepler} satellite to uniformly determine precise fundamental stellar parameters, including ages, for a sample of exoplanet host-stars where high-quality asteroseismic data were available. We devised a Bayesian procedure flexible in its input and applied it to different grids of models to study systematics from input physics and extract statistically robust properties for all stars. Our main results can be summarised as follows: 

\begin{description}
\item The stellar properties determined from each grid of models reflect the impact of the input physics in asteroseismic studies. The inclusion of microscopic diffusion clearly affects the ages derived while providing better fits to the spectroscopically determined effective temperature and metallicity. Extending the convective core size by means of overshooting results in slightly less massive stars than the standard Schwarzschild case once the convective core is well developed and the target is close to the TAMS. Interestingly, we show that including this effect favours a particular solution when bimodal distributions in mass are present.

\item Our results show that statistical uncertainties are below 2\% in radius and density, $\sim$3\% in mass, $\sim$4.5\% in distance and $\sim$14\% in age. This level of precision is perfectly suited for supporting characterisation of exoplanet systems, including determination of eccentricities by comparison of stellar densities from transit measurements with asteroseismic ones \citep[e.g.,][]{FogtmannSchulz:2014jd,VanEylen:2014cy}. Furthermore, our catalogue of stellar ages can be used for studying time evolution of planetary systems, tidal interactions and circularisation timescales. 

\item We have tested the systematic uncertainties introduced by changing the solar abundances and the efficiency of convection, and found them to be smaller than the statistical errors. We explored the impact on the stellar properties of reproducing different sets of asteroseismic observables (individual frequencies or frequency combinations), using different evolutionary and pulsation codes, as well as fitting algorithms, and found that when the input physics is kept the same the systematic differences are of the order of 1\% in density and radius, and 2\% and 9\% in mass and age. A source of uncertainty comparable to the statistical ones comes from the relation between metallicity and initial helium ($Y_{\rm ini}$), and we expect the \emph{Gaia} results to produce determinations of luminosity and provide an independent constraint on the initial helium abundances. Fitting algorithms that do not restrict this parameter commonly predict too low values of the $Y_{\rm ini}$, even below standard big bang nucleosynthesis. In Kepler-21, where a parallax measurement is available, the obtained luminosity of the sub-SBBN model is $1.5\,\sigma$ away from the parallax inferred value.

\item When the SNR of the data is too low and individual frequencies of oscillation are not possible to extract, asteroseismic determination of stellar properties is made using the global asteroseismic parameters. We have applied the BASTA using these observables and the spectroscopic constraints as input and found an excellent agreement with properties determined from frequency combinations. These results validate the use of the global asteroseismic observables in the regime explored in this paper when the theoretical value of $\langle\dnu\rangle$ is computed from individual frequencies of oscillations. If determined from the scaling relation instead, the results should be treated with caution as they seem to overestimate the density and underestimate the mass and radius.

\item Our results further demonstrate the positive synergy between asteroseismology and exoplanets studies, with the added ingredient of precise age determinations to a level better than 15\%. We have investigated age correlations with planetary period and radii in our sample, finding no noticeable trend in the results. Their distribution as a function of age appears constant and is likely representative of the underlying population of {\it Kepler} Sun-like host stars. In terms of resonances, our findings suggest that the period ratios in multiple systems tend to be near first order resonances regardless of the age of the system. Although subject to selection biases, our sample shows that the majority of planet host stars with asteroseismic observations of high SNR are older than the Sun.
\end{description}

The results presented here will provide the basis of upcoming catalogues of asteroseismic properties of exoplanet host-stars, where over 100 KOIs with $\langle\dnu\rangle$ and $\num$ measurements will be characterised (Huber et al. 2015, in preparation). This larger sample will extend the investigation we have initiated in this paper on the dependance of planetary properties with the age of the host star. Similarly, the BASTA combined with grids of evolutionary models will provide stellar properties for targets observed by the TESS and PLATO missions in evolutionary stages beyond the main-sequence phase.

\section*{Acknowledgements} 
Funding for the Stellar Astrophysics Centre is provided by The Danish National Research Foundation (Grant agreement No.~DNRF106). The research is supported by the ASTERISK project (ASTERoseismic Investigations with SONG and {\it Kepler}) funded by the European Research Council (Grant agreement No.~267864). G.R.D and W.J.C. acknowledges the support of the UK Science and Technology Facilities Council (STFC). S.B. acknowledges partial support of NSF grant AST-1105930 and NASA grant NNX13AE70G. T.S.M. was supported by NASA grant NNX13AE91G. Computational time on Kraken at the National Institute of Computational Sciences was provided through XSEDE allocation TG-AST090107. D.H. acknowledges support by the Australian Research Council's Discovery Projects funding scheme (project number DE140101364) and support by the National Aeronautics and Space Administration under Grant NNX14AB92G issued through the Kepler Participating Scientist Program. A.M.S acknowledges support from the grants ESP-2013-41268-R (MINECO) and 2014SGR-1458. The research leading to the presented results has received funding from the European Research Council under the European Community's Seventh Framework Programme (FP7/2007-2013) / ERC grant agreement no 338251 (StellarAges). C.K. acknowledges support from the Villum foundation.
\begin{landscape}
\clearpage 
\begin{table}\scriptsize
\caption{Recommended set of stellar properties and statistical uncertainties of exoplanet candidate host stars as determined with BASTA. Solar luminosity used $L_\odot=3.846\times10^{33}$~(erg$/$s). Additional systematic uncertainties can be accounted for (see sections~\ref{ssec_phys},~\ref{ssec_meth}, and~\ref{ssec_helium}): from input physics of 0.8\% (density), 0.7\% (radius), 2.3\% (mass), 9.6\% (age); from choice of observables 0.3\% (density and radius), 1\% (mass), and 7\% (age); from fitting algorithms and codes of 1\% (density and radius), 2\% (mass), and 9\% (age); from the initial helium abundance of 1.7\% (density), 1.6\% (radius), 3.6\% (mass), and 16.8\% (age). Reference column gives literature sources of confirmed or validated exoplanets. $^{\rm a}$see also \citet{Fressin:2011es}. $^{\rm b}$Ephemeris match indicates contamination, see CFOP website (https://cfop.ipac.caltech.edu/home/)}
\label{tab:stellar}
\begin{tabular}{ccccccccccccc}
\hline
\multicolumn{4}{ |c| }{} & \multicolumn{7}{ |c| }{BASTA} & &\\
\cline{5-11}
KOI & KIC & $\teff$ & $\feh$ & Mass & Radius & Density & $\logg$ & Luminosity & Age & Distance & Notes & Reference \\
 &  & (K) & (dex) &($\msun$) & ($R_\odot$) & (g$/$cm$^{3}$) & (dex) & ($L_\odot$) & (Gyr) & (pc) & & \\
\hline
\smallskip
   2 &        10666592 & 6350$\pm$ 80 &  0.26$\pm$0.08 & 1.497$^{+ 0.042}_{-0.040}$& 1.986$^{+ 0.018}_{-0.015}$& 0.269$^{+ 0.004}_{-0.005}$& 4.017$^{+ 0.007}_{-0.009}$&  5.754$^{+ 0.304}_{-0.322}$&  2.11$^{+ 0.29}_{-0.24}$&386.44$^{+ 12.11}_{- 11.95}$ &         HAT-P7 & \citet{2008ApJ...680.1450P}  \\
\smallskip
   5 &         8554498 & 5945$\pm$ 60 &  0.17$\pm$0.05 & 1.197$^{+ 0.021}_{-0.029}$& 1.794$^{+ 0.015}_{-0.015}$& 0.292$^{+ 0.003}_{-0.003}$& 4.007$^{+ 0.003}_{-0.003}$&  3.684$^{+ 0.130}_{-0.182}$&  5.60$^{+ 0.45}_{-0.42}$&439.34$^{+ 13.68}_{- 13.68}$ &                                           &   \\
\smallskip
   7 &        11853905 & 5781$\pm$ 76 &  0.09$\pm$0.10 & 1.117$^{+ 0.021}_{-0.029}$& 1.555$^{+ 0.012}_{-0.012}$& 0.419$^{+ 0.004}_{-0.006}$& 4.102$^{+ 0.005}_{-0.004}$&  2.505$^{+ 0.142}_{-0.124}$&  6.71$^{+ 0.77}_{-0.67}$&499.10$^{+ 15.46}_{- 15.46}$ &        Kepler-4 &\citet{2010Sci...327..977B}  \\
\smallskip
  41 &         6521045 & 5825$\pm$ 75 &  0.02$\pm$0.10 & 1.108$^{+ 0.021}_{-0.019}$& 1.513$^{+ 0.009}_{-0.012}$& 0.454$^{+ 0.004}_{-0.006}$& 4.125$^{+ 0.004}_{-0.004}$&  2.478$^{+ 0.112}_{-0.112}$&  6.50$^{+ 0.46}_{-0.56}$&310.34$^{+  9.49}_{-  9.63}$ &             Kepler-100 &\citet{Marcy:2014hr}  \\
\smallskip
  42 &         8866102 & 6325$\pm$ 75 &  0.01$\pm$0.10 & 1.228$^{+ 0.042}_{-0.040}$& 1.357$^{+ 0.012}_{-0.015}$& 0.695$^{+ 0.010}_{-0.011}$& 4.262$^{+ 0.007}_{-0.008}$&  2.753$^{+ 0.130}_{-0.130}$&  2.60$^{+ 0.56}_{-0.53}$&140.83$^{+  4.40}_{-  4.50}$ &        Kepler-410 A &\citet{VanEylen:2014cy}  \\
\smallskip
  69 &         3544595 & 5669$\pm$ 75 & -0.18$\pm$0.10 & 0.899$^{+ 0.009}_{-0.009}$& 0.916$^{+ 0.006}_{-0.006}$& 1.653$^{+ 0.016}_{-0.018}$& 4.468$^{+ 0.003}_{-0.003}$&  0.780$^{+ 0.029}_{-0.025}$&  6.63$^{+ 0.62}_{-0.57}$& 92.77$^{+  2.85}_{-  2.85}$ &            Kepler-93 &\citet{Ballard:2014je}  \\
\smallskip
  72 &        11904151 & 5647$\pm$ 74 & -0.15$\pm$0.10 & 0.918$^{+ 0.011}_{-0.019}$& 1.066$^{+ 0.006}_{-0.009}$& 1.068$^{+ 0.007}_{-0.012}$& 4.344$^{+ 0.003}_{-0.003}$&  1.063$^{+ 0.032}_{-0.034}$& 10.23$^{+ 0.83}_{-0.67}$&179.01$^{+  5.46}_{-  5.58}$ &  Kepler-10 &\citet{Batalha:2011fs}$^{\rm a}$  \\
\smallskip
  85 &         5866724 & 6169$\pm$ 50 &  0.09$\pm$0.08 & 1.199$^{+ 0.029}_{-0.030}$& 1.399$^{+ 0.012}_{-0.012}$& 0.616$^{+ 0.007}_{-0.008}$& 4.224$^{+ 0.005}_{-0.007}$&  2.577$^{+ 0.103}_{-0.095}$&  3.89$^{+ 0.59}_{-0.48}$&312.43$^{+  9.75}_{-  9.75}$ &            Kepler-65 &\citet{Chaplin:2013dg}  \\
\smallskip
 108 &         4914423 & 5845$\pm$ 88 &  0.07$\pm$0.11 & 1.099$^{+ 0.019}_{-0.030}$& 1.450$^{+ 0.009}_{-0.009}$& 0.507$^{+ 0.005}_{-0.005}$& 4.155$^{+ 0.004}_{-0.004}$&  2.256$^{+ 0.115}_{-0.119}$&  6.67$^{+ 0.69}_{-0.62}$&473.56$^{+ 14.51}_{- 14.51}$ &             Kepler-103 &\citet{Marcy:2014hr}  \\
\smallskip
 122 &         8349582 & 5699$\pm$ 74 &  0.30$\pm$0.10 & 1.068$^{+ 0.021}_{-0.019}$& 1.417$^{+ 0.012}_{-0.009}$& 0.528$^{+ 0.006}_{-0.005}$& 4.163$^{+ 0.003}_{-0.004}$&  1.891$^{+ 0.090}_{-0.094}$&  8.03$^{+ 0.80}_{-0.70}$&413.66$^{+ 12.89}_{- 12.68}$ &              Kepler-95 &\citet{Marcy:2014hr}  \\
\smallskip
 123 &         5094751 & 5952$\pm$ 75 & -0.08$\pm$0.10 & 1.068$^{+ 0.040}_{-0.040}$& 1.339$^{+ 0.015}_{-0.018}$& 0.628$^{+ 0.008}_{-0.007}$& 4.213$^{+ 0.008}_{-0.007}$&  2.040$^{+ 0.124}_{-0.112}$&  6.35$^{+ 1.05}_{-1.05}$&464.73$^{+ 14.88}_{- 15.28}$ &             Kepler-109 &\citet{Marcy:2014hr}  \\
\smallskip
 244 &         4349452 & 6270$\pm$ 79 & -0.04$\pm$0.10 & 1.159$^{+ 0.040}_{-0.051}$& 1.297$^{+ 0.015}_{-0.015}$& 0.745$^{+ 0.009}_{-0.010}$& 4.275$^{+ 0.007}_{-0.008}$&  2.406$^{+ 0.126}_{-0.128}$&  3.45$^{+ 0.81}_{-0.72}$&247.49$^{+  7.96}_{-  7.96}$ &            Kepler-25 &\citet{Steffen:2012kx}  \\
\smallskip
 245 &         8478994 & 5417$\pm$ 75 & -0.32$\pm$0.07 & 0.810$^{+ 0.019}_{-0.011}$& 0.772$^{+ 0.003}_{-0.006}$& 2.486$^{+ 0.022}_{-0.025}$& 4.570$^{+ 0.003}_{-0.002}$&  0.467$^{+ 0.023}_{-0.025}$&  5.35$^{+ 0.91}_{-1.01}$& 63.96$^{+  1.93}_{-  1.98}$ &           Kepler-37 & \citet{Barclay:2013fk}  \\
\smallskip
 246 &        11295426 & 5793$\pm$ 74 &  0.12$\pm$0.07 & 1.068$^{+ 0.011}_{-0.019}$& 1.237$^{+ 0.006}_{-0.006}$& 0.795$^{+ 0.005}_{-0.011}$& 4.280$^{+ 0.003}_{-0.003}$&  1.581$^{+ 0.043}_{-0.040}$&  6.31$^{+ 0.32}_{-0.34}$&139.77$^{+  4.25}_{-  4.25}$ &          Kepler-68 &\citet{Gilliland:2013ep}  \\
\smallskip
 260 &         8292840 & 6239$\pm$ 94 & -0.14$\pm$0.10 & 1.148$^{+ 0.051}_{-0.049}$& 1.345$^{+ 0.015}_{-0.018}$& 0.666$^{+ 0.010}_{-0.010}$& 4.240$^{+ 0.008}_{-0.008}$&  2.636$^{+ 0.148}_{-0.142}$&  3.85$^{+ 0.81}_{-0.75}$&240.28$^{+  7.69}_{-  7.89}$ &             Kepler-126  &\citet{Rowe:2014jq}  \\
\smallskip
 262 &        11807274 & 6225$\pm$ 75 & -0.00$\pm$0.08 & 1.239$^{+ 0.040}_{-0.040}$& 1.582$^{+ 0.015}_{-0.021}$& 0.444$^{+ 0.007}_{-0.007}$& 4.135$^{+ 0.007}_{-0.009}$&  3.466$^{+ 0.167}_{-0.157}$&  3.59$^{+ 0.78}_{-0.45}$&251.15$^{+  7.90}_{-  8.24}$ &           Kepler-50 & \citet{Steffen:2013ja}  \\
\smallskip
 263 &        10514430 & 5784$\pm$ 98 & -0.11$\pm$0.11 & 1.059$^{+ 0.040}_{-0.021}$& 1.590$^{+ 0.015}_{-0.018}$& 0.374$^{+ 0.004}_{-0.004}$& 4.061$^{+ 0.004}_{-0.004}$&  2.634$^{+ 0.151}_{-0.124}$&  7.84$^{+ 0.40}_{-0.91}$&211.53$^{+  6.65}_{-  6.78}$ &                   False-positive$^{\rm b}$ &  \\
\smallskip
 268 &         3425851 & 6343$\pm$ 85 & -0.04$\pm$0.10 & 1.178$^{+ 0.049}_{-0.049}$& 1.360$^{+ 0.015}_{-0.018}$& 0.662$^{+ 0.009}_{-0.008}$& 4.243$^{+ 0.008}_{-0.008}$&  2.728$^{+ 0.162}_{-0.157}$&  3.32$^{+ 0.85}_{-0.64}$&244.81$^{+  7.82}_{-  8.03}$ &                                           &   \\
\smallskip
 269 &         7670943 & 6463$\pm$110 &  0.09$\pm$0.11 & 1.239$^{+ 0.040}_{-0.051}$& 1.417$^{+ 0.015}_{-0.021}$& 0.616$^{+ 0.009}_{-0.010}$& 4.228$^{+ 0.008}_{-0.008}$&  3.039$^{+ 0.167}_{-0.173}$&  2.78$^{+ 0.62}_{-0.51}$&331.99$^{+ 10.56}_{- 11.11}$ &                                           &   \\
\smallskip
 274 &         8077137 & 6072$\pm$ 75 & -0.09$\pm$0.10 & 1.117$^{+ 0.040}_{-0.049}$& 1.632$^{+ 0.018}_{-0.009}$& 0.361$^{+ 0.006}_{-0.010}$& 4.056$^{+ 0.013}_{-0.010}$&  3.183$^{+ 0.293}_{-0.202}$&  6.23$^{+ 0.56}_{-1.23}$&413.44$^{+ 13.21}_{- 12.61}$ &              Kepler-128  &\citet{Xie:2014jk}  \\
\smallskip
 275 &        10586004 & 5770$\pm$ 83 &  0.29$\pm$0.10 & 1.178$^{+ 0.021}_{-0.030}$& 1.653$^{+ 0.009}_{-0.012}$& 0.367$^{+ 0.004}_{-0.005}$& 4.071$^{+ 0.005}_{-0.005}$&  2.749$^{+ 0.151}_{-0.155}$&  6.43$^{+ 0.64}_{-0.61}$&402.73$^{+ 12.28}_{- 12.43}$ &              Kepler-129 &\citet{Rowe:2014jq}  \\
\smallskip
 276 &        11133306 & 5982$\pm$ 82 & -0.02$\pm$0.10 & 1.059$^{+ 0.040}_{-0.030}$& 1.189$^{+ 0.012}_{-0.012}$& 0.891$^{+ 0.008}_{-0.009}$& 4.314$^{+ 0.004}_{-0.007}$&  1.628$^{+ 0.092}_{-0.079}$&  5.14$^{+ 0.86}_{-0.88}$&331.57$^{+ 10.50}_{- 10.50}$ &                                           &   \\
\smallskip
 277 &        11401755 & 5911$\pm$ 66 & -0.20$\pm$0.06 & 1.059$^{+ 0.030}_{-0.021}$& 1.632$^{+ 0.015}_{-0.021}$& 0.346$^{+ 0.003}_{-0.006}$& 4.039$^{+ 0.004}_{-0.004}$&  2.982$^{+ 0.144}_{-0.148}$&  7.10$^{+ 0.61}_{-0.59}$&514.89$^{+ 16.15}_{- 16.81}$ &             Kepler-36 &\citet{Carter:2012gq}  \\
\smallskip
 280 &         4141376 & 6134$\pm$ 91 & -0.24$\pm$0.10 & 1.019$^{+ 0.021}_{-0.030}$& 1.039$^{+ 0.009}_{-0.009}$& 1.278$^{+ 0.011}_{-0.013}$& 4.412$^{+ 0.003}_{-0.004}$&  1.363$^{+ 0.052}_{-0.056}$&  3.27$^{+ 0.59}_{-0.64}$&222.15$^{+  6.94}_{-  6.94}$ &                                           &   \\
\smallskip
 281 &         4143755 & 5622$\pm$106 & -0.40$\pm$0.11 & 0.918$^{+ 0.021}_{-0.030}$& 1.414$^{+ 0.012}_{-0.013}$& 0.456$^{+ 0.006}_{-0.006}$& 4.102$^{+ 0.001}_{-0.002}$&  1.997$^{+ 0.029}_{-0.038}$& 11.27$^{+ 1.50}_{-1.35}$&360.19$^{+ 11.23}_{- 11.30}$ &                                           &   \\
\smallskip
 285 &         6196457 & 5871$\pm$ 94 &  0.17$\pm$0.11 & 1.209$^{+ 0.019}_{-0.030}$& 1.716$^{+ 0.012}_{-0.012}$& 0.335$^{+ 0.004}_{-0.004}$& 4.049$^{+ 0.004}_{-0.005}$&  3.316$^{+ 0.185}_{-0.164}$&  5.52$^{+ 0.51}_{-0.48}$&448.57$^{+ 13.82}_{- 13.82}$ &                Kepler-92 &\citet{Xie:2014jk}  \\
\smallskip
 288 &         9592705 & 6174$\pm$ 92 &  0.22$\pm$0.10 & 1.509$^{+ 0.029}_{-0.021}$& 2.124$^{+ 0.015}_{-0.012}$& 0.221$^{+ 0.003}_{-0.003}$& 3.961$^{+ 0.004}_{-0.003}$&  6.088$^{+ 0.227}_{-0.247}$&  2.33$^{+ 0.18}_{-0.16}$&428.89$^{+ 13.22}_{- 13.09}$ &                                           &   \\
\smallskip
 370 &         8494142 & 6144$\pm$106 &  0.13$\pm$0.10 & 1.418$^{+ 0.030}_{-0.019}$& 1.887$^{+ 0.012}_{-0.012}$& 0.298$^{+ 0.005}_{-0.004}$& 4.038$^{+ 0.005}_{-0.005}$&  4.818$^{+ 0.259}_{-0.243}$&  2.62$^{+ 0.26}_{-0.24}$&584.10$^{+ 17.91}_{- 17.91}$ &               Kepler-145 &\citet{Xie:2014jk}  \\
\smallskip
 974 &         9414417 & 6253$\pm$ 75 & -0.13$\pm$0.10 & 1.387$^{+ 0.021}_{-0.009}$& 1.914$^{+ 0.015}_{-0.015}$& 0.280$^{+ 0.004}_{-0.005}$& 4.017$^{+ 0.004}_{-0.004}$&  5.343$^{+ 0.157}_{-0.157}$&  2.65$^{+ 0.14}_{-0.16}$&232.81$^{+  7.22}_{-  7.22}$ &                                           &   \\
\smallskip
 975 &         3632418 & 6305$\pm$ 50 & -0.03$\pm$0.10 & 1.408$^{+ 0.021}_{-0.030}$& 1.902$^{+ 0.018}_{-0.012}$& 0.287$^{+ 0.004}_{-0.005}$& 4.026$^{+ 0.004}_{-0.004}$&  5.188$^{+ 0.142}_{-0.128}$&  2.60$^{+ 0.16}_{-0.16}$&114.49$^{+  3.60}_{-  3.51}$ &             Kepler-21 &\citet{Howell:2012jv}  \\
\smallskip
1612 &        10963065 & 6104$\pm$ 74 & -0.20$\pm$0.10 & 1.089$^{+ 0.019}_{-0.030}$& 1.228$^{+ 0.012}_{-0.009}$& 0.823$^{+ 0.010}_{-0.009}$& 4.293$^{+ 0.004}_{-0.003}$&  1.961$^{+ 0.058}_{-0.058}$&  4.18$^{+ 0.53}_{-0.35}$& 92.37$^{+  2.91}_{-  2.85}$ &             Kepler-408 &\citet{Marcy:2014hr}  \\
\smallskip
1925 &         9955598 & 5460$\pm$ 75 &  0.08$\pm$0.10 & 0.890$^{+ 0.009}_{-0.011}$& 0.883$^{+ 0.003}_{-0.006}$& 1.821$^{+ 0.018}_{-0.018}$& 4.495$^{+ 0.002}_{-0.002}$&  0.598$^{+ 0.025}_{-0.023}$&  6.98$^{+ 0.40}_{-0.50}$& 66.80$^{+  2.02}_{-  2.05}$ &             Kepler-409 &\citet{Marcy:2014hr}  \\
\smallskip
3158 &         6278762 & 5046$\pm$ 74 & -0.37$\pm$0.09 & 0.739$^{+ 0.009}_{-0.011}$& 0.748$^{+ 0.006}_{-0.003}$& 2.498$^{+ 0.018}_{-0.025}$& 4.560$^{+ 0.003}_{-0.002}$&  0.341$^{+ 0.020}_{-0.016}$& 11.54$^{+ 0.99}_{-0.94}$& 32.90$^{+  1.02}_{-  1.00}$ &         Kepler-444 & \citet{Campante:2015ei}  \\
\hline
\end{tabular}
\end{table}
\end{landscape}
\footnotesize{
\bibliographystyle{mn2e}
\bibliography{kages}
}
\appendix
\section{Evolutionary codes and fitting algorithms}\label{app_meth}
In the following sections we describe the fitting procedures used to determine stellar properties for our sample, including the evolutionary and pulsation codes used.
\subsection{ASTFIT}\label{ssec_fit_jcd}
The ASTEC Fitting method (ASTFIT) uses the Aarhus stellar evolution \citep[ASTEC,][]{ChristensenDalsgaard:2008bi} and pulsation \citep[ADIPLS,][]{ChristensenDalsgaard:2008kr} codes. It is based on fitting individual model frequencies, applying a correction for the surface errors, to the observed frequencies \citep[see also][]{Gilliland:2013ep}. The fit is carried out in grid of models varying the mass and composition. The chemical composition is based on a helium enrichment relation $\Delta Y/\Delta Z = 1.4$, set up in terms of initial hydrogen and heavy-element abundances $(X_0, Z_0)$, with three neighbouring values of $Z_0$ for each value of $X_0$. The grid extends in $\feh$ between $-0.577$ and 0.415 in steps of around 0.05 and in mass between $0.7$ and 1.69~$\msun$, in steps of 0.01~$\msun$. The ASTFIT results presented here were computed without diffusion and settling, and convective-core overshoot was not considered. The mixing-length parameter was set to $\alpha_{\rm MLT} = 1.8$, roughly corresponding to solar calibration. The model physics otherwise essentially followed the BASTA specifications.

The fit of a given model to the data is defined in terms of 
\begin{equation}
\chi^2_\nu = {1 \over N-1} \sum_{i=1}^N 
\left( {\nu^{\rm (obs)}_i - \nu^{\rm (mod)}_i \over \sigma_i } \right)^2 \; ,
\label{eq:chisqnu}
\end{equation}
where $\nu^{\rm (obs)}_i$ and $\nu^{\rm (mod)}_i$ are the observed and model frequencies and $\sigma_i$ is the error in the observed frequencies. Here the model frequencies typically include a surface correction (see section~\ref{sec:surf} below). It is assumed that the degree and order of the observed frequencies have already been determined, and that they do not include additional mixed modes. Hence, when the models contain mixed modes, they are excluded (renumbering the remaining modes) retaining only the most acoustically dominated modes (as determined from their inertia) in each interval in large frequency separation $\dnu$. The fit also includes
\begin{equation}
\chi_{\rm spec}^2 = 
\left( {T_{\rm eff}^{\rm (obs)}- T_{\rm eff}^{\rm (mod)} 
\over \sigma(T_{\rm eff}) } \right)^2  
+ \left( {{\rm [Fe/H]}^{\rm (obs)}- {\rm [Fe/H]}^{\rm (mod)} 
\over \sigma({\rm [Fe/H]}) } \right)^2  \; ,
\label{eq:chisqspec}
\end{equation}
with the observed and model effective temperature $\teff$ and $\feh$. The complete fit is then characterised by $\chi^2 = \chi^2_\nu + \chi^2_{\rm spec}$, defining the relative weight between the oscillation and the spectroscopic data.

For each evolution sequence frequencies are calculated for at least every second timestep. Amongst these frequency sets the model $\CM_{\rm min}^\prime$ with the smallest $\chi_\nu^2$ is determined. Based on homology scaling it is assumed that the frequencies in the vicinity of $\CM_{\rm min}^\prime$ can be obtained as $r \nu_i(\CM_{\rm min}^\prime)$ where $r = [R/R(\CM_{\rm min}^\prime)]^{-3/2}$, $R$ being the surface radius of the model,
and the best-fitting model is determined by minimising
\begin{equation}
\chi^2_\nu(r) = {1 \over N-1} \sum_{i=1}^N 
\left( {\nu^{\rm (obs)}_i - r \nu_i(\CM_{\rm min}^\prime) 
\over \sigma_i } \right)^2
\label{eq:chisqfit}
\end{equation}
as a function of $r$. The resulting value $r_{\rm min}$ of $r$ defines an estimate $R_{\rm min}$ of the radius of the best-fitting model along the given sequence. 

The other properties of this best-fitting model, for the given set of model parameters $\{\CP_k\}$ and with corresponding $\chi_{\nu,\rm min}^2(\CP_k)$ are determined by linear interpolation in $R$ to $R_{\rm min}$. The corresponding minimum $\chi_{\rm min}^2(\CP_k)$ is then determined as $\chi_{\rm min}^2(\CP_k) = \chi_{\nu, \rm min}^2(\CP_k) 
+ \chi_{\rm spec}(\CM_{\rm min}^\prime)$\footnote{and hence neglecting the generally negligible change in, e.g., $T_{\rm eff}$ between timesteps.}. The final best-fitting model is found as the parameter set corresponding to the smallest $\chi_{\nu,\rm min}^2(\CP_k)$ (or $\chi_{\rm min}^2(\CP_k)$) over a suitable selection of models in the grid.

It is very informative to plot $\chi_{\nu,\rm min}^2(\CP_k)$ or $\chi_{\rm min}^2(\CP_k)$ against the various model properties, such as mean density $\langle \rho \rangle$, $R$ or age. Representative results are given in the HAT-P-7 paper \citep{ChristensenDalsgaard:2010hx}. The best-fitting frequencies, e.g., for comparison with the observations in an \'echelle diagram, are obtained by applying the appropriate scaling $r_{\rm min}$ to the frequencies of the model $\CM_{\rm min}^\prime$ in the minimising sequence.

Although probably not completely justified in a statistical sense, estimates of stellar properties, in particular radius, mass and age, are determined as averages and standard deviations of the properties of the models $\CM_{\min}(\CP_k)$ over the parameters $\{\CP_k\}$, with the weights $\exp(-\chi_{\rm min}^2(\CP_k)/2)$, corresponding to the likelihood. These are the final ASTFIT results compared with the BASTA results in figures~\ref{fig:fitting_rho} and~\ref{fig:fitting_mass} and presented in Table\ref{tab:astfit}.
\subsubsection{Near-surface correction}\label{sec:surf}
It is well established that errors in the modelling of the structure of the near-surface layers of the star and their effects on the oscillations cause systematic differences between the model and observed frequencies, even when the model otherwise matches the structure of the star. This is particularly evident in the solar case, where the availability of frequencies over a broad range of degrees allow separation between the effects of the bulk of the solar interior and the surface layers. These errors must be taken into account when, as done in ASTFIT, individual model frequencies are fitted to the observations. The correction unavoidably requires prior and perhaps poorly justified assumptions, typically based on the behaviour determined in the solar case.

\begin{figure}
\includegraphics[width=84mm]{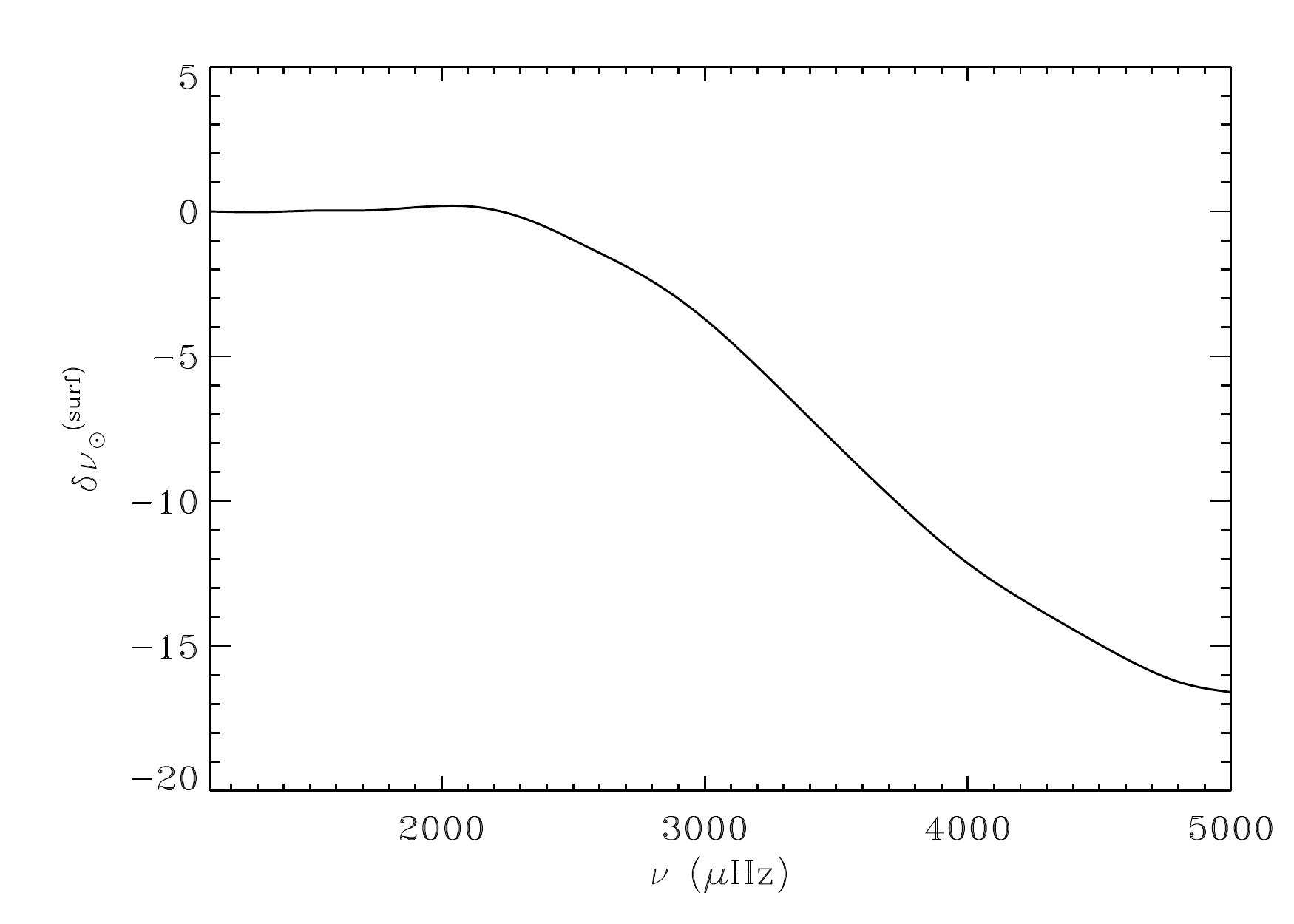}
\caption{The solar surface correction $\delta_\odot^{\rm (surf)}$
determined from
$\CH_2$ in the fit given in Eq. (\ref{eq:asfit}) to solar data over
a broad range of degrees.}
\label{fig:solsurf}
\end{figure}

Here we specifically use a representation of the surface correction that has a functional form matching the solar behaviour \citep[see also][]{Aerts:2010uw,ChristensenDalsgaard:2012bg}. This is obtained from the differential form of the Duvall asymptotic expression for the frequencies \citep{1989MNRAS.238..481C}, according to which the frequency differences $\delta \nu$ between the Sun and a model satisfy
\begin{equation}
S_{nl} {\delta \nu_{nl} \over \nu_{nl}} \simeq
\CH_1(\nu_{nl}/L) + \CH_2(\nu_{nl}) \; ,
\label{eq:asfit}
\end{equation}
where $L = \sqrt{l(l+1)}$ and $S_{nl}$ is a scale factor which can be calculated from the model, and which may be chosen to be close to 1 for low-degree modes. The term in $\CH_1$ reflects the difference in sound speed between the Sun and the model throughout the star, whereas $\CH_2$ contains the contribution from the near-surface region.
Since on general grounds the surface effects are very small at low frequency, the arbitrary additive constant in the definition of $\CH_2$ may be chosen such that $\CH_2$ is zero at low frequency. In Fig.~\ref{fig:solsurf} is shown the frequency shift $\delta \nu_\odot^{\rm (surf)}$ caused by the surface effects in the Sun, determined in this manner from $\nu \CH_2(\nu)$.
We now make the assumption that the correction in other stars has a similar functional form, in terms of frequency measured in units the acoustic cut-off frequency $\nu_{\rm ac}$, which provides a physically motivated frequency scale for the stellar atmosphere. Thus for each set of model frequencies we obtain the surface correction by determining a scale factor $r$ and an amplitude $\tilde a$ from a least-squares fit of
\begin{equation}
\nu_{nl} = r \nuref + \tilde a \CG_\odot(\nu_{nl}/\nu_{\rm ac}) 
\label{eq:solsurf}
\end{equation}
to the observed frequencies, where the function $\CG_\odot$ is determined from $\nu \CH_2(\nu)$, illustrated in Fig.~\ref{fig:solsurf}, using a solar acoustic cut-off frequency of
$\nu_{\rm ac,\odot} = 5200~\mu$Hz. The model frequencies in equation~\ref{eq:chisqnu} are obtained from this expression, but without including the scale factor.
\begin{table*}\scriptsize
\caption{Stellar properties determined with ASTFIT. Solar luminosity used $L_\odot=3.846\times10^{33}$~(erg$/$s). The first five objects are shown here for guidance on the format, and the full table is available online.}
\label{tab:astfit}
\begin{tabular}{cccccccc}
\hline
KOI & KIC & Mass & Radius & Density & $\logg$ & Luminosity & Age  \\
 &  & ($\msun$) & ($R_\odot$) & (g$/$cm$^{3}$) & (dex) & ($L_\odot$) & (Gyr) \\
\hline
\smallskip
   2 &        10666592 & 1.538$\pm$ 0.030 & 2.000$\pm$ 0.012 & 0.271$\pm$ 0.002 & 4.023$\pm$ 0.004 & 5.811$\pm$ 0.213 &  1.88$\pm$ 0.14   \\
\smallskip
   5 &         8554498 & 1.173$\pm$ 0.020 & 1.785$\pm$ 0.010 & 0.291$\pm$ 0.001 & 4.004$\pm$ 0.003 & 3.373$\pm$ 0.133 &  6.12$\pm$ 0.28   \\
\smallskip
   7 &        11853905 & 1.118$\pm$ 0.031 & 1.553$\pm$ 0.013 & 0.420$\pm$ 0.002 & 4.104$\pm$ 0.005 & 2.488$\pm$ 0.117 &  7.27$\pm$ 0.69   \\
\smallskip
  41 &         6521045 & 1.121$\pm$ 0.012 & 1.517$\pm$ 0.006 & 0.452$\pm$ 0.001 & 4.125$\pm$ 0.002 & 2.509$\pm$ 0.088 &  6.83$\pm$ 0.28   \\
\smallskip
  42 &         8866102 & 1.218$\pm$ 0.036 & 1.354$\pm$ 0.013 & 0.691$\pm$ 0.004 & 4.260$\pm$ 0.005 & 2.654$\pm$ 0.111 &  2.78$\pm$ 0.47 \\
\hline
\end{tabular}
\end{table*}
\subsection{YMCM}\label{ssec_fit_bas}
The YALE Monte-Carlo Method used the YREC \citep[][]{Demarque:2007ij} code to model the stars. While the input physics was the same as in the other techniques, the method of constructing the models was very different.

The starting point of the modelling effort for each star was the average large separation that was used in the grid-based modelling, the mass obtained from grid-based modelling and the effective temperature and metallicity obtained from spectroscopy. Since each of these quantities is associated with an uncertainty, we created many more realisations of these parameters to obtain a larger set of ($M$, $\dnu$, $\teff$, $\feh$). The $\dnu$ in each set, along with $M$ was used to calculate $R$ using the usual scaling relationship. Since the scaling relationship is not exact, we used the empirical correction of \citet{White:2011fw} to get a better estimate.

For each combination ($M$, $R$, $\teff$, $\feh$) we used YREC in an iterative mode to obtain a model of the given mass $M$ and $\feh$ than had the required $R$ and $\teff$.  Since the mixing length parameter was set to be the solar calibrated value ($1.6756$ for the given physics and formulation of mixing-length theory in YREC), the only free parameter we had was the initial helium abundance $Y_0$ that was varied until the required model was obtained. Some combinations of parameters resulted in models that need $Y_0$ to be lower than the primordial helium abundance; all such models were rejected. 

Theoretical frequencies of oscillation are computed using the code first described by \citet{1994A&AS..107..421A}. The $\chi^2$ for the differences between the observed frequencies and those of the models, $\chi^2_\nu$, were calculated after correcting for the surface term in a manner outlined below. Similarly, we calculated $\chi^2_{\rm rat}$ for the differences in the frequency ratios $\runo$ and $\rdos$ (including correlations), $\chi^2_{\teff}$ for the difference in effective temperature, and $\chi^2_{\feh}$ for the difference in metallicity. The final goodness-of-fit was determined by
\begin{equation}
\chi^2_{\rm tot}=\chi^2_\nu+\chi^2_{\rm rat}+\chi^2_{\teff}+\chi^2_{\feh}.
\end{equation}
The best-fit model was the one with the lowest value of $\chi^2_{\rm tot}$.

We used the scaled solar surface term to correct the frequency differences between the observed and model frequencies. We first constructed a solar model with the same physics as the rest of the models and determined its frequency differences with respect to observed solar frequencies obtained by the Birmingham Solar Oscillations Network (BiSON) and listed in Table~1 of \citet{Chaplin:2007bg}. The solar surface term was obtained by fitting these frequency differences to the differential form of the Duvall Law \citep[][]{1989MNRAS.238..481C}. We denote this as the $\nu_{nl,\odot}$-$\delta\nu_{nl,\odot}$ relation. Both  $\nu_{nl,\odot}$ and $\delta\nu_{nl,\odot}$ are then scaled to the mass and radius of the stellar model under consideration using the homology scaling 
\begin{equation}
r=\frac{\langle\Delta\nu({\rm mod})\rangle}{\langle\Delta\nu_\odot\rangle},
\end{equation}
where the angular brackets denote the average. The resulting $r\nu_{nl,\odot}$-$r\delta\nu_{nl,\odot}$  is then used to correct the stellar model for the surface term. The corrected frequencies are denoted as
\begin{equation} 
\nu_{nl}^{\rm (corr)}=\nu_{nl}^{\rm (mod)}+\beta r\delta\nu_{nl,\odot},
\label{eq:my_sur} 
\end{equation} 
with $r\delta\nu_{nl,\odot}$ evaluated at $r\nu_{nl,\odot}$=$\nu_{nl}({\rm obs})$. The factor $\beta$ is selected to minimize
\begin{equation}
\sum\frac{\left(\nu_{nl}^{\rm(obs)}-\nu_{nl}^{\rm(corr)}\right)^2}{\sigma^2(\nu_{nl})}.
\end{equation}
$\nu_{nl}^{\rm (corr)}$ are then used to calculate $\chi^2_\nu$. The final set of stellar properties determined with the YMCM are available in Table~\ref{tab:ymcm}.
\begin{table*}\scriptsize
\caption{Stellar properties determined with YMCM. Solar luminosity used $L_\odot=3.846\times10^{33}$~(erg$/$s). The first five objects are shown here for guidance on the format, and the full table is available online.}
\label{tab:ymcm}
\begin{tabular}{cccccccccc}
\hline
KOI & KIC & Mass & Radius & Density & $\logg$ & Luminosity & Age & $Y_{\rm ini}$ & $Z_{\rm ini}$ \\
 &  & ($\msun$) & ($R_\odot$) & (g$/$cm$^{3}$) & (dex) & ($L_\odot$) & (Gyr) & & \\
\hline
\smallskip
   2 &        10666592 & 1.546$\pm$ 0.035 & 2.001$\pm$ 0.016 & 0.272$\pm$ 0.002 & 4.024$\pm$ 0.004 & 5.744$\pm$ 0.175 &  1.63$\pm$ 0.10 &0.278$\pm$0.017 & 0.0282$\pm$0.0038   \\
\smallskip
   7 &        11853905 & 1.146$\pm$ 0.030 & 1.567$\pm$ 0.014 & 0.419$\pm$ 0.005 & 4.107$\pm$ 0.005 & 2.533$\pm$ 0.100 &  6.93$\pm$ 0.55 &0.257$\pm$0.013 & 0.0232$\pm$0.0026   \\
\smallskip
  41 &         6521045 & 1.138$\pm$ 0.009 & 1.525$\pm$ 0.004 & 0.452$\pm$ 0.001 & 4.127$\pm$ 0.001 & 2.564$\pm$ 0.048 &  6.22$\pm$ 0.14 &0.267$\pm$0.006 & 0.0219$\pm$0.0020   \\
\smallskip
  42 &         8866102 & 1.238$\pm$ 0.041 & 1.359$\pm$ 0.015 & 0.694$\pm$ 0.003 & 4.264$\pm$ 0.005 & 2.671$\pm$ 0.083 &  2.32$\pm$ 0.31 &0.268$\pm$0.016 & 0.0187$\pm$0.0021   \\
\smallskip
  69 &         3544595 & 0.919$\pm$ 0.020 & 0.922$\pm$ 0.007 & 1.650$\pm$ 0.002 & 4.471$\pm$ 0.003 & 0.771$\pm$ 0.028 &  7.04$\pm$ 0.68 &0.252$\pm$0.015 & 0.0133$\pm$0.0015   \\
\hline
\end{tabular}
\end{table*}
\subsection{AMP}\label{ssec_fit_tra}
For the results presented in section~\ref{ssec_meth}, we used the Asteroseismic Modeling Portal \citep[AMP,][]{Metcalfe:2009ed, 2009gcew.procE...1W} in a similar configuration as that described in \cite{Metcalfe:2014ig}, but without including diffusion and settling of helium. In summary, AMP uses a parallel genetic algorithm \citep[GA,][]{Metcalfe:2003ka} to optimize the match between stellar model output and the available set of observational constraints. The evolution models are produced with ASTEC, and the oscillation frequencies are calculated with ADIPLS. The five adjustable model parameters include the mass ($M$), age ($t$), composition ($Z$ and $Y_\mathrm{i}$), and mixing-length ($\alpha$). The oscillation frequencies and other properties of each model are compared to four sets of observational constraints, including: [1] the individual frequencies corrected for surface effects following the empirical prescription of \cite{Kjeldsen:2008kw}, [2] the frequency ratios $\rdos$ defined by Eq.~\ref{eqn:r02}, [3] the frequency ratios $\runo$ defined by Eq.~\ref{eqn:r010}, and [4] the available spectroscopic constraints. A normalized $\chi^2$ is calculated for each set of constraints, and the GA attempts to minimize the mean of the four $\chi^2$ values. This allows the various asteroseismic quality metrics to be traded off against each other, while ensuring that the numerous frequencies and ratios do not overwhelm the relatively few spectroscopic constraints. The interested reader can find the optimal model parameters produced by AMP for each star in Table~\ref{tab:amp}.
\begin{table*}\scriptsize
\caption{Stellar properties determined with AMP. Solar luminosity used $L_\odot=3.846\times10^{33}$~(erg$/$s). The first five objects are shown here for guidance on the format, and the full table is available online.}
\label{tab:amp}
\begin{tabular}{ccccccccccc}
\hline
KOI & KIC & Mass & Radius & Density & $\logg$ & Luminosity & Age & $Y_{\rm ini}$ & $Z_{\rm ini}$ & $\alpha_{\rm MLT}$ \\
 &  & ($\msun$) & ($R_\odot$) & (g$/$cm$^{3}$) & (dex) & ($L_\odot$) & (Gyr) & & & \\
\hline
\smallskip
   2 &        10666592 & 1.300$\pm$ 0.055 & 1.879$\pm$ 0.018 & 0.276$\pm$ 0.001 & 4.004$\pm$ 0.003 & 5.096$\pm$ 0.291 &  2.79$\pm$ 0.52 &0.317$\pm$0.011 & 0.0239$\pm$0.0043 &1.72$\pm$0.13  \\
\smallskip
   5 &         8554498 & 1.310$\pm$ 0.038 & 1.825$\pm$ 0.018 & 0.303$\pm$ 0.001 & 4.032$\pm$ 0.005 & 3.444$\pm$ 0.190 &  4.56$\pm$ 0.60 &0.243$\pm$0.011 & 0.0255$\pm$0.0032 &1.44$\pm$0.09  \\
\smallskip
   7 &        11853905 & 1.140$\pm$ 0.017 & 1.553$\pm$ 0.016 & 0.428$\pm$ 0.008 & 4.112$\pm$ 0.007 & 2.435$\pm$ 0.185 &  6.13$\pm$ 0.30 &0.268$\pm$0.012 & 0.0217$\pm$0.0027 &1.62$\pm$0.12  \\
\smallskip
  41 &         6521045 & 1.090$\pm$ 0.018 & 1.501$\pm$ 0.013 & 0.454$\pm$ 0.008 & 4.123$\pm$ 0.005 & 2.337$\pm$ 0.129 &  7.16$\pm$ 0.42 &0.265$\pm$0.010 & 0.0184$\pm$0.0012 &1.78$\pm$0.07  \\
\smallskip
  42 &         8866102 & 1.260$\pm$ 0.032 & 1.363$\pm$ 0.014 & 0.701$\pm$ 0.004 & 4.269$\pm$ 0.003 & 2.626$\pm$ 0.137 &  2.01$\pm$ 0.47 &0.257$\pm$0.012 & 0.0184$\pm$0.0015 &1.82$\pm$0.09 \\
\hline
\end{tabular}
\end{table*}
\end{document}